\documentclass[aps,prb,superscriptaddress,showpacs,citeautoscript,twocolumn]{revtex4-1}

\usepackage{graphicx}
\usepackage{amsmath}
\usepackage{amssymb}
\usepackage{color}

\newcommand{\bea}{\begin{eqnarray*}}
\newcommand{\eea}{\end{eqnarray*}}
\newcommand{\bne}{\begin{equation*}}
\newcommand{\ede}{\end{equation*}}

\newcommand{\bnen}{\begin{equation}}
\newcommand{\eden}{\end{equation}}
\newcommand{\bean}{\begin{eqnarray}}
\newcommand{\eean}{\end{eqnarray}}
\newcommand{\bnsn}{\begin{subequations}}

\newcommand{\edsn}{\end{subequations}}

\newcommand{\bna}{\begin{array}}
\newcommand{\eda}{\end{array}}
\newcommand{\bnm}{\begin{enumerate}}
\newcommand{\edm}{\end{enumerate}}

\newcommand{\spinor}[2]{\left(\bna{c} #1 \\[1.5ex] #2 \eda\right)}

\newcommand{\ttmatrix}[4]{\left(\bna{cc} #1 & #2 \\ #3 & #4 \eda\right)}

\renewcommand{\vec}[1]{\text{\boldmath{$ #1 $}}}

\newcommand{\bra}[1]{\langle #1 |}
\newcommand{\ket}[1]{| #1 \rangle}

\begin{document}
\title{Probing individual split Cooper-pairs using the spin qubit toolkit}

\author{Zolt\'an Scher\"ubl}
\affiliation{Department of Physics, Budapest University of Technology and Economics and
Condensed Matter Research Group of the Hungarian Academy of Sciences, 1111 Budapest, Budafoki \'ut 8., Hungary}
\author{Andr\'as P\'alyi}
\affiliation{Department of Materials Physics, E\"otv\"os University, P\'azm\'any P\'eter s\'et\'any 1/A, H-1117 Budapest, Hungary}
\affiliation{BME-MTA Exotic Quantum Phases Research Group,
Budapest University of Technology and Economics, Budapest,
Hungary}
\author{Szabolcs Csonka}
\affiliation{Department of Physics, Budapest University of Technology and Economics and
Condensed Matter Research Group of the Hungarian Academy of Sciences, 1111 Budapest, Budafoki \'ut 8., Hungary}

\date{\today}

\begin{abstract}
A superconductor is a natural source of spin-entangled spatially separated electron pairs. Although the first Cooper-pair splitter devices have been realized recently, an experimental confirmation of the spin state and the entanglement of the emitted electron pairs is lacking up to now. In this paper a method is proposed to confirm the spin-singlet character of individual split Cooper pairs. Two quantum dots (QDs), each of them holding one spin-prepared electron, serve as the detector of the spin state of a single Cooper pair that is forced to split when it tunnels out from the superconductor to the QDs. The number of charges on the QDs, measured at the end of the procedure, carries information on the spin state of the extracted Cooper pair. The method relies on the experimentally established toolkit of QD-based spin qubits: resonant spin manipulation, Pauli blockade, and charge measurement.
\end{abstract}


\maketitle

\section{Introduction}

Generation and control over  entangled quantum states is a first step towards the development of future quantum machines. The electron spin is a promising candidate to represent quantum information in such systems \cite{LossPRA1998}. A superconductor is a natural source of spin entangled electron pairs, since in the BCS ground state electrons form Cooper-pairs, which are entangled spin singlet pairs.

With the extraction of individual Cooper-pairs and separation of the consisting electrons to two normal leads, two streams of mobile entangled electrons could be generated \cite{LesovikEPJB2001}. This principle is implemented in the so-called Cooper-pair splitter device (CPS), which contains quantum dots (QDs) at the interface of the superconductor (SC) and the two normal leads \cite{RecherPRB2001}. Due to Coulomb repulsion on the dots, the two electrons of a Cooper-pair cannot enter the same dot, thereby the desired spatial separation of the electron pairs can be achieved.

The original scheme of Recher et al. \cite{RecherPRB2001} motivated  intensive theoretical\cite{ChtchelkatchevPRB2002,Eldridge-cps,Hiltscher-cps,ChevallierPRB2011,RechPRB2012,CottetPRL2012,Leijnse-spinqubits,VeldhorstPRL2010, BednorzPRB2011, BursetPRB2011, HilscherPRB2012,SatoPRB2012, SollerBeilstein2012, CottetPRB2012,ChenPRB2013, GiovannettiPRB2012} and experimental \cite{HofstetterNature2009,HermannPRL2010,HerrmannArXiv2012,HofstetterPRL2011,DasNatComm2012,WeiNatPhys2010,SchindelePRL2012,LambertArXiv2013} efforts to analyze the Cooper-pair splitting process. The first CPS devices were fabricated very recently based on semiconductor nanowires (NWs)\cite{HofstetterNature2009,HerrmannArXiv2012} and carbon nanotubes (CNTs) \cite{HermannPRL2010}. The Cooper-pair splitting process was analysed at finite bias condition \cite{HofstetterPRL2011} and was demonstrated even in current cross correlation \cite{DasNatComm2012,WeiNatPhys2010}. Furthermore, splitting efficiency up to 90\% has also been  demonstrated \cite{SchindelePRL2012}.

So far the performed measurements focused on the charge correlation of the two outputs of the CPS device. The natural next step is to address the spin character and the level of entanglement of the spatially separated electron pairs. Theoretical proposals exist  for such tests, like adding ferromagnetic detectors \cite{ChtchelkatchevPRB2002} at the outputs of the CPS, combining it with a beam mixer unit \cite{BurkardPRB2000} or place the CPS in a cavity \cite{CottetPRL2012}. However their experimental realization is quite challenging, since e.g. the first scheme requires highly spin polarized and rotatable  ferromagnetic contacts, while the other two are based on demanding sample geometry.

In this work, we propose a novel way to confirm the spin-singlet character of individual split Cooper-pairs based on the toolkit of spin qubits \cite{HansonRMP2007}, i.e., on experimental techniques developed in the past decade to coherently manipulate and read out localized electronic spins in solids.
In our proposed experiment, the spin character is tested directly on the two QDs of the CPS. First, one electron is placed in each QD, and their spins
are prepared in known quantum states. Then a Cooper pair is forced to split from the superconductor to the QDs. Due to Pauli's exclusion principle, the probability of a successful splitting event is determined by the spin state of the prepared electrons as well as on the spin state of the split Cooper-pair. This probability can be measured by charge readout on the QDs at the end of the procedure. By performing this measurement for various initial spin states of the QD electrons, the spin-singlet character of the split Cooper pairs can be confirmed. The building blocks of the proposed scheme were all demonstrated before, therefore our proposal can be realized with state-of-the-art experimental techniques.

\section{Setup}

The proposed device geometry is shown in Fig.~1. From the normal CPS geometry we focus on the SC electrode and two neighboring QDs, $L$ and $R$.
The charge occupation of each QD can be measured by a nearby charge sensor (CS). The CSs can be realized by, e.g., quantum point contacts or additional QDs, which are capacitively coupled to QDs $L$ and $R$ \cite{FieldPRL1993}.
In the present proposal, the tunneling from the QDs to the normal leads (N) is switched off and the normal leads are not used. 
The level positions of the QDs can be manipulated with the voltages 
of the gate electrodes (G). 
Independent manipulation of the spins residing in the two dots can be 
performed via electrically driven spin resonance 
(EDSR) \cite{FlindtPRL2006,GolovachPRB2006}, using extra local
gates (not shown).
(Alternatively, the plunger gates G themselves can be used to 
control the spins.\cite{NadjPergeNature2010,SchroerPRL2011})


EDSR is an important ingredient in the proposed experiment outlined in Sec. \ref{sec:proposal}. This mechanism of coherent single-spin control has been  experimentally demonstrated, among other systems, in QDs in semiconductor nanowires\cite{NadjPergeNature2010,SchroerPRL2011,BergArXiv2012,PribiagArXiv2013} and carbon nanotubes\cite{PeiNatureNano2012,LairdNatureNano2013}, both being important platforms for Cooper-pair splitters. In these materials, coherent Rabi oscillations with Rabi frequencies up to 100 MHz, corresponding to spin-flop times of the order of $10$ ns, have been measured under electrical excitation. As discussed below, local addressability of the spin qubits is required in our present proposal, which is relatively easily satisfied by EDSR where the spins are controlled via ac voltages applied to local gates. Besides the experimental advances, the theoretical understanding of the  microscopic mechanisms underlying EDSR in QDs is also developing rapidly \cite{FlindtPRL2006,GolovachPRB2006,BulaevPRL2007,RashbaPRB2008,RuiLiPRL2013,KloeffelArXiv2013,FlensbergPRB2010,SzechenyiArXiv2013} It seems certain that in semiconductors, a strong spin-orbit interaction is beneficial for fast EDSR. As a combined effect of the electrical drive and spin-orbit interaction, the spin qubit feels an effective ac magnetic field  $\hbar \boldsymbol\Omega(t)$ that induces Rabi oscillations. The orientation of the effective ac field might be linked to a certain crystallographic direction of the crystal lattice, but it can also be influenced by the sample design and the electrostatic potential landscape used to form the QD. \cite{CsonkaNanoLett2008,SchroerPRL2011}

In summary, the ingredients of the proposed device geometry, like coupling QDs on the two sides of a SC \cite{HofstetterNature2009, HermannPRL2010}, performing EDSR on the spin state of QDs \cite{NowackScience2007,NadjPergeNature2010} and readout the charge state of the QDs with CS \cite{HuNatureNano2012,KuemmethMatToday2010} have been all demonstrated in semiconductor nanowires or carbon nanotubes based devices.

\begin{figure}[!htbp]
	\begin{center}
	\includegraphics[width=8.5cm]{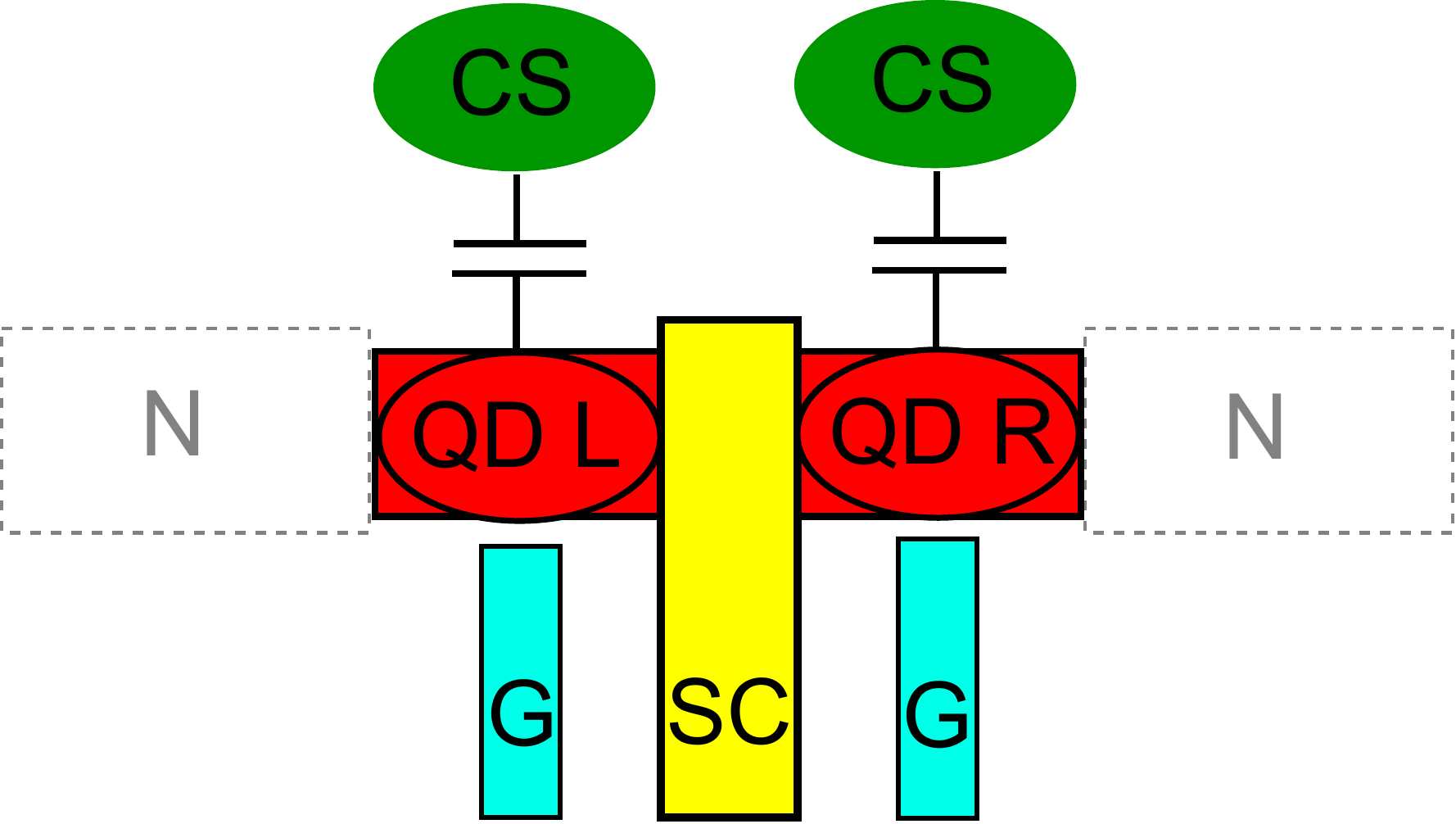}
	\caption{The suggested Cooper-pair splitter geometry. The main part is the superconducting electrode (SC) tunnel coupled to two, separated quantum dots (QD) $L$ and $R$. The energy levels of the electrons on QD $L$ and $R$ are manipulated with voltage of the gate (G) electrodes. 
	Independent manipulation of the electron spins can be performed via electrically driven spin resonance using extra local gates (not shown).The charge state of the dots is measured by capacitively coupled charge sensors (CS). Note that normal leads (N, dashed) 
	of the usual Cooper-pair splitter geometry 
	have no role in the proposed measurement.}
	\label{fig1}
	\end{center}
	\end{figure}
	
\section{The model}
\label{sec:themodel}

For the sake of simplicity, we consider the case when the occupation of a single orbital level is allowed in both QDs (see Fig.~2c), and the on-site energies $\epsilon_L$ and $\epsilon_R$ of the two QDs are controlled simultaneously,  $\epsilon_L = \epsilon_R = \epsilon$. Coulomb interaction between electrons in different QDs is effectively screened by the superconductor in between, and therefore we disregard it. On-site electron-electron
interaction is taken into account via the Coulomb energy $U$. The energy scale characterizing the tunneling between each QD and the SC is the tunnel amplitude $t$, see also Appendix \ref{sec:derivation}. We consider the weak-tunneling regime:
\bnen t \ll \Delta, U, \eden
where $\Delta$ is the energy gap of the superconductor.

In the experiment proposed below, the dynamics is essentially restricted to the states with no quasiparticles in SC, and QD charge configurations $(0,0)$, $(1,1)$ and $(2,2)$. Here $(n,m)$ denotes the class of states where QD $L$ ($R$) is occupied by $n$ ($m$) electrons. States with other charge configurations and states including a finite number of quasiparticles
in SC might be involved in the dynamics only perturbatively. In the absence of SC-QD tunneling, the energies of the $(0,0)$, $(1,1)$ and $(2,2)$ charge configurations of the two dots are $0$, $2\epsilon$, and $4 \epsilon +2U$, respectively.

In the presence of weak SC-QD tunneling, transitions via virtual intermediate states, consisting of an odd number of electrons in the two QDs and a single quasiparticle in the superconductor, induce coherent coupling between different even-electron charge configurations of the QDs. The coupling is especially effective between the $(0,0)$ and $(1,1)$ [$(2,2)$ and $(1,1)$] charge states in the vicinity of $\epsilon = 0$ [$\epsilon = -U$], where the energies of the $(0,0)$ and the $(1,1)$ [$(2,2)$ and $(1,1)$] charge states would
coincide in the absence of SC-QD tunneling. Using the BCS Hamiltonian for the superconductor, and assuming spin-conserving and left-right symmetric SC-QD
tunneling [see \eqref{eq:Htunnel}],
we derive effective Hamiltonians for the QD states from quasi-degenerate perturbation theory. For the case of the $(0,0)-(1,1)$ anticrossing at $\epsilon \approx 0$, we find
\bean \label{eq:heff0} H_{\rm eff} (\epsilon \approx 0) &=& 2\epsilon \sum_{\sigma} \ket{\sigma (1,1)}\bra{\sigma (1,1)} \nonumber \\ &+& \left(\tilde \Delta \ket{S(1,1)}\bra{(0,0)} + \mbox{h.~c.} \right) \eean
where the sum is for the four spin states of the $(1,1)$ charge configuration,  $\sigma \in (S,T_+,T_0,T_-)$. The value of the coupling parameter $\tilde \Delta$ and the validity of the perturbative treatment depend on the geometry of the device, the SC-QD tunnel amplitudes, the superconducting gap, and the band structure of the superconductor (see Appendix \ref{sec:derivation}
for details). The effective Hamiltonian at the $(1,1)-(2,2)$ anticrossing at $\epsilon \approx -U$ reads
\bean \label{eq:heffmU} H_{\rm eff} (\epsilon \approx -U) &=& 2\epsilon \sum_{\sigma}\ket{\sigma (1,1)}\bra{\sigma (1,1)} \nonumber \\ &+& (4\epsilon + 2U) \ket{(2,2)}\bra{(2,2)} \nonumber \\ &-& \left( \tilde \Delta \ket{S(1,1)}\bra{(2,2)} + \mbox{h.~c.} \right) \eean
Note that energy shifts of second order in the SC-QD tunneling, determining the precise position of the anticrossings, as well as
higher-order terms in the SC-QD tunneling strength $t$,
are omitted from the above effective Hamiltonians.

The states of the $(1,1)$ charge configuration are sensitive to the presence of real or effective magnetic  fields. These interactions are described by the Zeeman Hamiltonian
\bnen \label{eq:spinham} H_m = \sum_{D\in \{L,R\}} 
\left[ \vec{\mathcal{B}}_D + \hbar \boldsymbol \Omega_D(t)\right] \cdot \vec S_D, 
\eden
where $\vec{\mathcal{B}}_D$ is the time-independent field
and 
$\hbar \boldsymbol \Omega_D(t)$ is the ac field, both 
having the dimension of energy,
and $\vec{S}_D$ is the spin vector operator of the electrons in QD $D$. The 
dc effective magnetic field $\vec{\mathcal{B}}_D$ incorporates the effects of the static external magnetic field $\vec B_{\rm ext}$ and the Overhauser-field $\vec{\mathcal{B}}_{N,D}$ induced by the nuclear spins residing in QD $ D$:
\bnen \label{eq:spinhamiltonian}
\vec{\mathcal{B}}_D = \mu_B \hat{g}_D \vec B_{\rm ext} + \vec{\mathcal{B}}_{N,D}, \eden
The g-tensors $\hat g_D$ might differ on the two dots. The ac 
effective magnetic field $\hbar \vec{\Omega}_D(t)$ can arise from, e.g., ac electrical excitation via EDSR\cite{NowackScience2007,NadjPergeNature2010,GolovachPRB2006,FlindtPRL2006}. For clarity, we first treat the simple, idealized case of $\hat g_L = \hat g_R$ and $\vec{\mathcal{B}}_{N,D} = 0$ in this Section and Sec. \ref{sec:proposal}, and then discuss the effect of deviations from this idealized case in Section \ref{sec:discussion}.

Fig.~2a summarizes the effect of the SC-QD tunnel coupling  on the QDs energy levels in the presence of a static external magnetic field. Due to the Zeeman effect, the $T_+$, $S,T_0$, and  $T_-$ levels split, while the S-QD tunnel coupling induces anticrossing of the $S(1,1)$ and $(0,0)$ (and $(2,2)$) states  at $\epsilon \approx 0$ ($\epsilon \approx -U$). The triplet subspace remains uncoupled to the $(0,0)-S(1,1)-(2,2)$ subspace.
The hybridization of the $S(1,1)$ and $(0,0)$ at $\epsilon \approx 0$ implies that if the two QDs are prepared in the $(0,0)$ charge configuration ($\epsilon > 0$) and the level positions of the QDs is lowered adiabatically e.g. to  $\epsilon \approx -U/2$ the system ends up in the $S(1,1)$ state.
This gate voltage sweep results in the extraction of a single Cooper pair from the superconductor. In the following, a measurement scheme
is described which uses the other anticrossing (at $\epsilon \approx -U$) to address the spin character of an individual split Cooper-pair.

\begin{figure*}[!htbp]
	\begin{center}
	\includegraphics[width=0.9\textwidth]{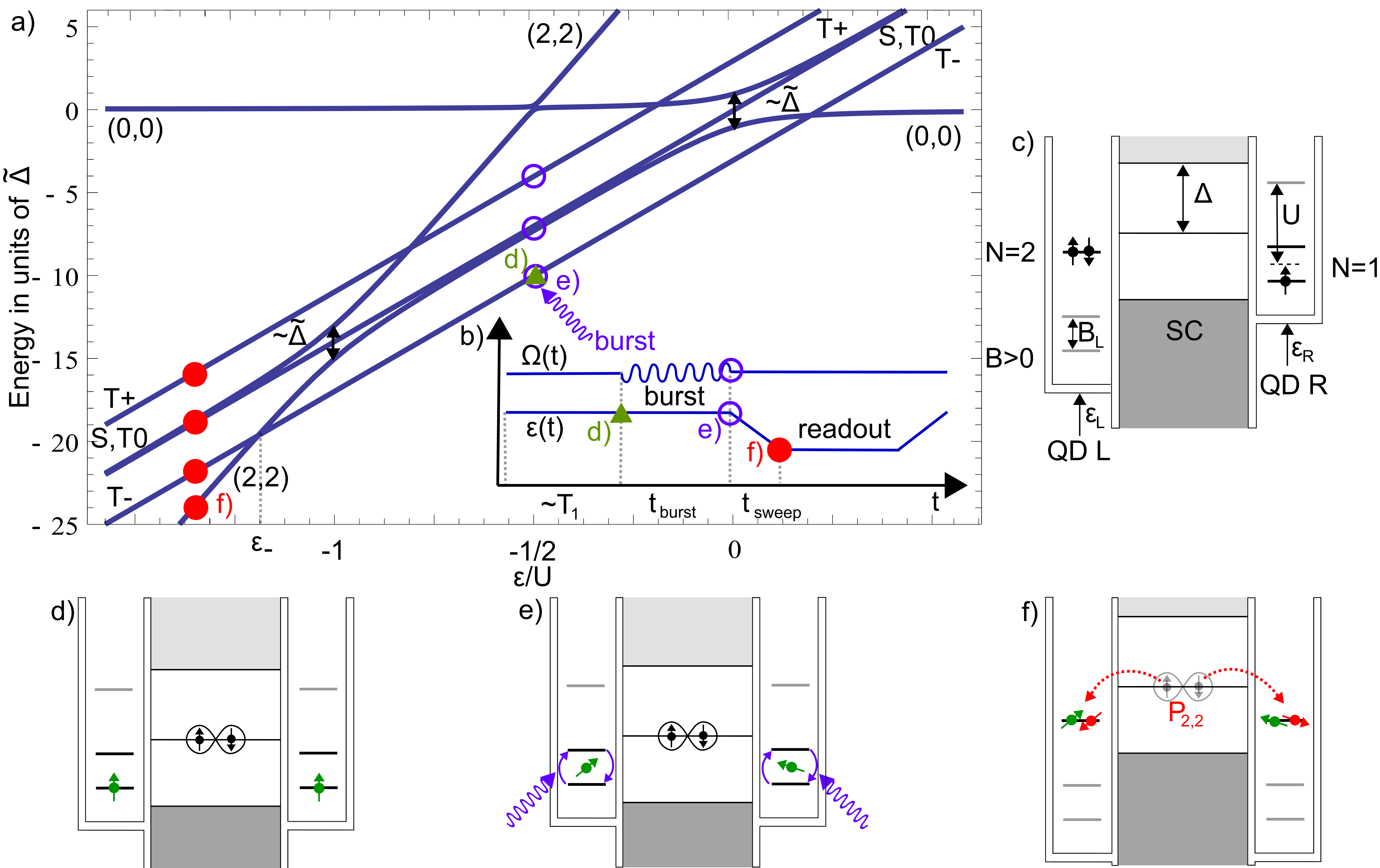}
	\caption{a) Energy spectrum of the tunnel coupled QD-SC-QD system as a function of the level position of the QDs, $\epsilon$, with $ \mu_B g B_{\rm ext} =1.5\tilde\Delta$ and $U=7\tilde\Delta$. The levels $T_+$ and $T_-$ are split from $S$ and $T_0$ due to the Zeeman effect induced by the external B field. Due to the tunnel-coupling at $\epsilon=0$ [$\epsilon=-U$], the states $(0,0)$ [$(2,2)$] and the singlet state $S$ hybridizes. b) Gate voltage sequence for the proposed detection scheme. c) Energy levels of the QDs and the SC electrode, where U is the charging energy, $\Delta$ is the SC gap, $B_D$ represents the Zeeman splitting, $\epsilon_D$ is the level position of QD D, $D=L,R$. Charge configuration $(2,1)$ is shown.
	d-f) Schematic representations of the steps of the measurement scheme. The corresponding points are marked on the a) and b) subfigures.}
	\label{fig2}
	\end{center}
	\end{figure*}

\section{The proposed experiment}
\label{sec:proposal}

In this Section, we outline the proposed experiment that allows the demonstration of the spin-singlet character of the Cooper-pairs extracted from the SC. Here we consider an idealized case where unwanted perturbations influencing spin dynamics are absent. Effects of such perturbations (such as
hyperfine interaction and the different g-tensors on the two QDs) are discussed in the subsequent Section.

Figs.~2d-f show the steps of the detection scheme. A finite, static magnetic field $\vec B_{\rm ext}$ induces a Zeeman spin splitting in both QDs.
As a starting point, the common on-site energy $\epsilon$ of the QDs is set at an initial position in the vicinity of $\epsilon = -U/2$, where the $S(1,1)$ and $T_0(1,1)$ are (approximately) degenerate. Waiting longer than the spin relaxation time, $T_1$ of the QDs, the system relaxes to the $T_-$ state (Fig.~2d).

Since the SC-QD tunnel coupling $t$ is weak compared to $U$ and $\Delta$, the level structure of the $(1,1)$ sector is almost unaffected by the tunnel coupling at $\epsilon \approx -U/2$ (see Fig.~2a). Therefore the spin state of the two QDs can be manipulated independently. By applying EDSR pulses on 
QDs $L$ and $R$,  an arbitrary non-entangled spin state of the two electrons can be prepared (see Fig.~2e).
This prepared state is denoted as
$\ket{\theta_L,\phi_L(t);\theta_R,\phi_R(t)}$,
where $\theta_D$ and $\phi_D$ are the polar and azimuth angles of the electron spin on QD $D$ on the Bloch sphere with respect to $B_{ext}$. Due to the Larmor-precession around the external field, this state evolves in time, but as the magnetic fields are the same on the two QDs, the speed of Larmor-precession is equal, hence the difference of the azimuth angles $\phi_L(t) - \phi_R(t)$ is steady in time.

In general, the state $\ket{\theta_L,\phi_L(t);\theta_R,\phi_R(t)}$ contains contribution from all four spin states of the $(1,1)$ charge configuration (defined after Eq.~2). This prepared state serves to detect the spin character of an individual split Cooper pair. Adiabatically lowering the level positions to $\epsilon < -U$, a Cooper pair tries to tunnel from the superconductor to the QDs. According to Fig.~2a the $(2,2)$ state hybridizes only with $S(1,1)$, therefore the Cooper pair can only leave the SC if the prepared $(1,1)$ state has singlet contribution. Thus at the end of the sequence, the probability $P_{2,2}$ to find the QDs in the $(2,2)$ charge configuration, that is, the probability of the successful tunneling event of the Cooper pair, is equal to
\bnen P_{2,2} = |\langle S(1,1) \mid \theta_L,\phi_L(t);\theta_R,\phi_R(t)\rangle|^2 \label{prob}. \eden
(Note that $P_{2,2}$ is not time dependent.) When a single sequence is finished, the charge state of the QDs is read out by the charge detectors, which show either the $(1,1)$ or the $(2,2)$ states. 
\footnote{Importantly, higher-lying QD orbitals should not be populated during the $\epsilon$ sweep before the charge measurement. This requirement is fulfilled if the energies of the orbitally excited (2,2) states exceed the energy of $T_+(1,1)$ at the measurement point $\epsilon = \epsilon_m$ (red points in Fig. 2a), which is the case if the orbital level spacing exceeds $2|U +  \epsilon_m| + 4\mathcal{B}_{\rm ext}$.} Then the QDs are set back to $\epsilon \approx -U/2$, and the whole sequence is repeated several times to determine $P_{2,2}$.

The result of Eq.~(6) can also be interpreted as a direct consequence of Pauli-exclusion principle and spin conservation during the tunneling events: In the $(2,2)$ state both $\uparrow$ and $\downarrow$ spin state are occupied on both QDs, thus the total spin of the four electrons is zero. The  magnitude of the spin of the extracted Cooper pair is also zero, therefore the QDs can absorb the Cooper pair only if
the prepared $(1,1)$ state has zero spin as well. The probability $P_{2,2}$ corresponds to those cases; otherwise,
the tunneling of the Cooper pair is blocked. This mechanism is similar to the conventional Pauli-blockade effect in double QD systems~\cite{Ono-Science2002}, however in the present case the spin state of \textit{two separated} QDs has to match with the spin state of \textit{two} outcoming electrons, thus the Pauli blockade has to be fulfilled simultanously on both QDs of the CPS.

As long as the ac effective magnetic field pulses $\vec \Omega_L(t)$ and $\vec \Omega_R(t)$ are parallel, in phase, and started synchronously, the relation
\bnen \label{eq:phase} \phi_L(t) = \phi_R(t) \eden
holds.
For this case, the probability $P_{2,2}$ to find the system in the $(2,2)$ charge state at the end of the measurement sequence is shown in Fig.~3. $P_{2,2}$ is plotted as a function of $\theta_L$ and $\theta_R$, i.e., the polar rotation angles of the EDSR pulses on the two QDs. The value of $P_{2,2}$ varies between 0 and $0.5$.  $P_{2,2}$ takes its maximum e.g. at $\theta_L=0$ and $\theta_R = \pi$, in this case prepared $(1,1)$ state is $\ket{\downarrow,\uparrow}$, on which a single Cooper-pair state can tunnel out with probability of $1/2$. $P_{2,2}$ has its minimum along the diagonal, i.e. when $\theta_L=\theta_R$. For these angles, the prepared spins on the two QDs are parallel to each other, therefore these states have a \textit{pure triplet} character. According to the simultaneous Pauli-blockade on the two QDs, a Cooper pair can only tunnel out from the SC if the prepared $(1,1)$ state has singlet character (see Eq.~6), which leads to $P_{2,2}$ equals zero along the diagonal. The zero probability along the diagonal line is a benchmark of the singlet character of the split Cooper-pair: Let us assume for a moment that the electron pairs coming from the middle 'superconductor' lead would have triplet contribution as well (e.g. either the middle lead is not a singlet source or the electron pair loses the singlet character during tunneling to the QDs). In this case the $T(1,1)$ states also hybridize with the $(2,2)$ charge state and thus the probability to extract an electron pair ($P_{2,2}$) would be finite for certain prepared pure $T(1,1)$ state as well. Thus $P_{2,2}=0$ would not hold along the entire diagonal.
 

In conclusion a spin sensitive manipulation sequence was outlined to analyze the spin character of split Cooper pairs. First the spin state of the QDs is prepared by EDRS, than the QDs energy levels are lowered adiabatically, finally the charge state of the dots is read out.  At the end of the 
sequence, the probability $P_{2,2}$ of finding both QDs doubly occupied,  as a function of the rotation angles $\theta_L$ and $\theta_R$ shows a characteristic pattern, which is a direct consequence of the singlet character of Cooper pairs. Therefore, performing the outlined manipulation sequence and evaluating $P_{2,2}(\theta_L,\theta_R)$ the singlet character of individual split Cooper pairs can be determined.

	\begin{figure}[!hbp]
	\begin{center}
	\includegraphics[width=9cm]{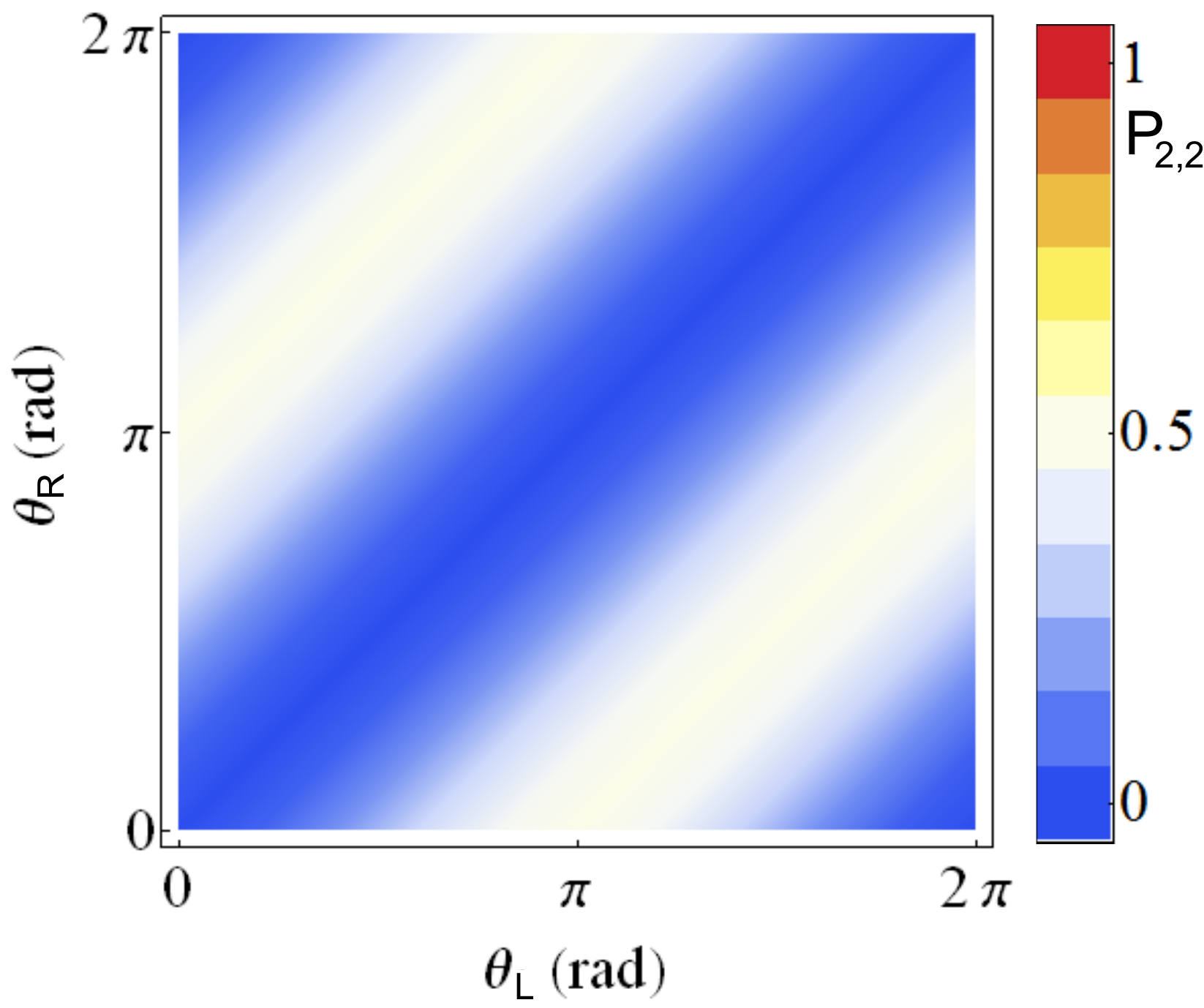}
	\caption{The probability $P_{2,2}$ of detecting $(2,2)$ charge configuration at the end of the manipulation sequence as a function of $\theta_L$ and $\theta_R$ polar rotation angles. The  azimuth angles $\phi_L$ and $\phi_R$ are assumed to be equal.}
	\label{fig3}
	\end{center}
	\end{figure}

\section{Discussion}
\label{sec:discussion}

In the previous Section, we discussed the proposed experiment in an idealized case with the following simplifications:
(A) nuclear spins are absent, $\vec{\mathcal{B}}_{N,D} = 0$,
(B) the g-factors are identical on the two QDs, $g_L = g_R$,
(C) the gate voltage sweep between the preparation, measurement
points is adiabatic,
(D) the ac effective magnetic fields are parallel, $\vec \Omega_L \parallel \vec  \Omega_R$, and
(E) the microwave pulses for spin control are started exactly at the same
time.
In a real experiment, at least some of these conditions are relaxed, potentially leading to important differences in the result with respect to the idealized case. In this Section, we discuss such differences: after a brief account of the role of (D) and (E), we discuss (A), (B), and (C) in detail.

(D) A well-suited mechanism for local spin control is the spin-orbit mediated EDSR, which allows for inducing coherent Rabi oscillations with ac voltage pulses on the gate electrodes defining the QDs. In practice, the directions of the effective ac fields $\vec \Omega_L$ and $\vec \Omega_R$ (see Eq.~\eqref{eq:spinham}) driving these Rabi oscillations depend on the electrostatic potential landscape of the QD as well as on the character of spin-orbit interaction in the material. As the two QDs in a CPS device are not necessarily identical, the directions of the corresponding effective ac fields might also differ.

Let us consider a specific example to illustrate the effect of different ac field directions, i.e., of $\vec \Omega_L \nparallel \vec \Omega_R$.
Assume the directions of $\vec \Omega_L$ and $\vec \Omega_R$ are known, and that the dc magnetic field vectors are the same on the two dots.
In the rotating frame, the spin rotation axis corresponding to the Rabi oscillation in each dot is determined by
(i) the direction of the projections of the ac field vectors to the plane transversal to the dc field, and
(ii) the phase of the microwave voltage pulse driving the spin rotation. If the phases of the microwaves are the same in the two dots, then the misalignment between the transversal projection of $\vec \Omega_L$ and $\vec \Omega_R$ implies misaligned Rabi rotation axes in the rotating frame, which translates to a finite relative phase difference of the Larmor precession of the two spins (in the lab frame). Hence Eq.~\eqref{eq:phase} does not hold, therefore the outcome of the measurement of $P_{2,2}$ will be different from the pattern shown in Fig.~3. However, since the misalignment angle of $\vec \Omega_L$ and $\vec \Omega_R$ is known, an appropriate phase difference in the microwave pulses can be applied in order to align the Rabi rotation axes of the two spins in the rotating frame, hence to bring the Larmor precession of the two spins back in phase (i.e., to restore Eq.~\eqref{eq:phase}), and thereby to allow for the observation of the pattern of $P_{2,2}$ shown in Fig.~3.

(E) Perfect timing of the spin-controlling microwave voltage pulses is probably impossible. If the typical random uncertainty in the start time of
the pulses is $\delta t$, then the typical phase lag of the Larmor precession of the two spins is $\delta \phi = g \mu_B B_{\rm ext} \delta t / \hbar$.
The condition $\delta \phi \ll 1$ should hold in order to observe the pattern of Fig.~3. For a g-factor of $g=2$ and magnetic field $B_{\rm ext} = 50$ mT, the latter condition approximately translates to $\delta t \ll 100$ ps. Note that the effect of a deterministic, reproducible  lag between the starting time of the pulses can be compensated by adjusting the phase of one of the pulses.

\subsection{Nuclear spins}
\label{sec:nuclearspins}

If the material hosting the QDs has nuclear spins, then hyperfine interaction is present, giving rise to two random and independent
effective magnetic fields (`Overhauser fields') for the electrons in the two QDs. The Overhauser field in QD $D$, in energy units, is denoted by $\vec{\mathcal{B}}_{N,D}$, see Eq.~\eqref{eq:spinhamiltonian}. Although these  fields average to zero, their standard deviations are finite and they induce different Zeeman-type splittings on the two dots with values of $\mathcal B_{N,D}$, and therefore they influence the corresponding Larmor precession frequencies.
Thus this random contribution of magnetic field causes a finite inhomogeneous spin dephasing time $T^*_2$, which is of the order of $10$ ns for InAs\cite{NadjPergeNature2010} and InSb\cite{BergArXiv2012} NW QDs .
Here we assume that the standard deviations of the Overhauser-field components in the two dots are identical. 
The standard deviation of the Overhauser-field component parallel to the external magnetic field, expressed in energy units, 
 is denoted by $\mathcal{B}_N$.
The latter quantity is related to the inhomogeneous dephasing time 
as\cite{HansonRMP2007} $T_2^* = \sqrt{2} \hbar / \mathcal{B}_N$.

Consider the case when, in our proposed experiment, the g-tensors are isotropic and equal, the EDSR drive frequency is set to the nominal resonance frequency ($\hbar \omega = g \mu_B B_{\rm ext}$), the rotating wave approximation holds 
($g \mu_B B_{\rm ext} \gg \hbar \Omega_L, \hbar \Omega_R$), and the EDSR Rabi frequency exceeds the hyperfine-induced Zeeman splitting ($\hbar \Omega_L, \hbar \Omega_R \gg \mathcal{B}_N$). The latter condition has two consequences. The first one is that the EDSR pulse induces complete Rabi oscillations for practically
any value of the Overhauser field; the second one is that the Overhauser field is unable to induce a significant Larmor-phase difference between the two spins during a Rabi cycle. Right after the spin manipulation is completed, a sufficiently fast sweep of $\epsilon$ towards the measurement point (red points in Fig. 2a) switches off the hyperfine-induced dephasing, hence the measurement result is expected to be close to the ideal case shown in of Fig. 3.

In a material with many nuclear spins, it is possible that the hyperfine-induced Zeeman splitting exceeds the EDSR Rabi frequency,
$\mathcal{B}_N \gg \hbar \Omega$. In this case, the resonance frequency is strongly shifted by the instantaneous value of the Overhauser field, therefore driving at the frequency matching the nominal Zeeman splitting ($\hbar \omega = g \mu_B B_{\rm ext}$)  is unlikely to cause Rabi oscillations. (Numerical results for $P_{2,2}$ and their explanations for the intermediate regime  $\mathcal{B}_N \sim \hbar \Omega_L, \hbar \Omega_R$ can be found in Appendix \ref{app:nuclearspins}) As a consequence, materials with weak hyperfine interaction, or devices with  large effective ac fields are preferred for our proposed experiment.

Taking the example of a semiconductor nanowire based n-type QD \cite{BergArXiv2012}, the manipulation time of a $2\pi$ rotation of $\theta$ is possible within $\sim 10$ ns. This time scale is comparable to the $T^*_2$ time, therefore the experimental observation of the main features of the pattern shown in Fig. 3 seems only feasible in III-V  NW devices if the dephasing time can be prolonged or the spin-flip time can be decreased.
Considering systems with weaker hyperfine interaction, such as hole-based QDs with $p$ type wave function or nuclear-spin free systems, such as isotopically purified Si/Ge nanowires or carbon based QDs, $T^*_2$ might be further increased \cite{Fischer-hole,Fischer-carbon}, potentially allowing for the observation of the ideal-case result of $P_{2,2}$ shown in Fig.~3.

\subsection{Different g-tensors on the two QDs}
\label{sec:disc2}

In typical semiconducting nanowire or carbon nanotube QDs, the g-tensor is anisotropic\cite{SchroerPRL2011,KuemmethNature2008}.
As the g-tensor can be strongly influenced by the local electrostatic 
potential landscape via spin-orbit coupling, the two g-tensors in a double QD (DQD)
might differ significantly.
Hence, in a general case, for a given $\vec B_{\rm ext}$, the magnitude and the direction of the effective fields $\vec{\mathcal{B}}_{D} = \mu_B \hat{g}_D \vec B_{\rm ext}$ are different on the two QDs. In the following, the expected outcome of the proposed experiment is discussed for two cases: a) when the effective fields are parallel, but their magnitudes are different; b) when  the magnitudes of the effective fields are the same, but their direction encloses an angle.

a) A large g-factor difference of the two dots usually implies different Zeeman splittings, making it necessary to independently tune the frequencies of the microwave pulses driving EDSR in the two QDs.

Furthermore the g-factor difference of the  QDs generates different Larmor precession. For instance taking a typical $t_{burst} \approx 5 ns$ and $g_L-g_R = 2$ at $B_{\rm ext}=50 mT$ a large phase difference $\Delta \phi = t_{burst} \mu_B (g_L - g_R) B_{\rm ext}/\hbar \approx 15\pi$ accumulates between the azimuthal angle of the two spins during the preparation.  A fix $\Delta \phi$ is not a problem for the proposed measurement sequence, since its influence can be taken into account upon calculating $P_{2,2}$. However even a small uncertainty of the pulse length smears the characteristic features of $P_{2,2}$. If the uncertainty of $\Delta \phi$ reaches $ \approx \pi$, 
then the relative weights of the $S$ and $T_0$ components
of the prepared (1,1) state become randomized.
Therefore the scheme loses its ability to identify the singlet
character of the Cooper pairs.
Accordingly one should try for reducing the difference of the g-factors.
 


	\begin{figure}[!hbp]
	\begin{center}
	\includegraphics[width=9cm]{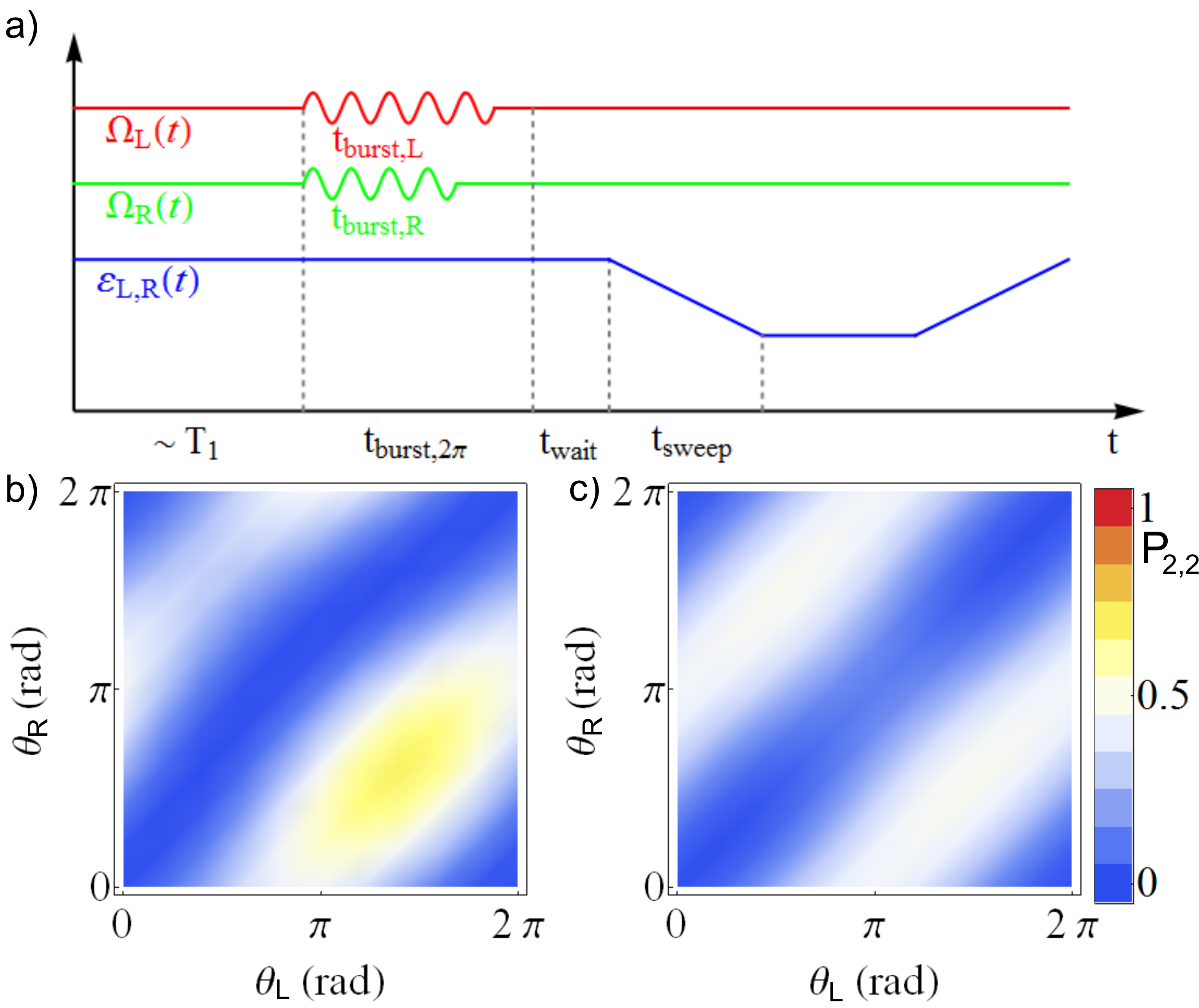}
	\caption{
	Different g-tensors in the two dots: pulse sequence
	and simulation results.
	a) Schematic representation of the 
	pulse sequence used in the simulation of the 
	proposed experiment.
	b,c) Simulation results 
	(for details, see Appendix \ref{app:gtensorsimulation}) 
	for the
	probability map $P_{2,2}(\theta_L,\theta_R)$, in the 
	case of different g-tensors in the two dots.
	Zeeman splittings in the two dots are equal, but there is a
	a finite angle $\beta=32^\circ$ enclosed by
	the effective dc magnetic fields in the two dots.
	b) $P_{2,2}$ map for $t_{\rm wait}=23$ ps.
	c) $P_{2,2}$ maps averaged for $t_{\rm wait}$ for one Larmor period,
	$t_{\rm wait} \in [0,42]$ ps.
	Results b) and c) should be compared to the ideal-case result of
	Fig. 3.}
	\label{fig4}
	\end{center}
	\end{figure}

b) 
The anisotropic nature of the g-tensors can help to reduce 
the unwanted difference of the Zeeman splittings on the two QDs. As described in Appendix \ref{app:gtensoranisotropic}, if the surfaces corresponding to the g-tensors of the two QDs have an intersection, the direction of the external magnetic field can be chosen so that the Zeeman-splitting is the same for the two QDs, i.e. 
$ |\vec{\mathcal{B}}_{L}|=|\vec{\mathcal{B}}_{R}|$.
For instance in the double-dot NW sample used in Ref.~ \onlinecite{SchroerPRL2011}, the Zeeman splittings in the two dots can be tuned equal (see intersection of surfaces in Fig. \ref{fig6}e).
In this situation, the same Larmor frequency is set  for the spins in the	two QDs, but the 
 Larmor precession takes place around the two different axes,
 defined by the directions of $\vec{\mathcal{B}}_L$ and 
 $\vec{\mathcal{B}}_R$, enclosing an angle $\beta$.	

Due to the different Larmor-precession axes, the angle between the spin polarization vectors of the two QDs changes 
periodically in time with the Larmor period.
This implies that the singlet component of the prepared spin state,
and hence
the measurement outcome $P_{2,2}$, will depend on the protocol
of the spin preparation, e.g., on the length and the strength of the applied
Rabi pulses. 
This is in contrast to the ideal-case scenario detailed in Sec. 
\ref{sec:proposal}, where $P_{2,2}$ depends only on the
spin rotation angles $\theta_L$ and $\theta_R$, and is insensitive
to any other detail of the spin manipulation protocol.

It is natural to expect that for $\beta \ll 1$, the $P_{2,2}$ probability
map obtained at the end of our scheme is very similar to
the ideal-case ($\beta = 0$) result shown in Fig. 3, irrespective
of the parameters specifying the Rabi pulses. 
Here, we use numerical simulation to demonstrate that even for a relatively
large angle, up to $\beta \lesssim \pi/6 \equiv 30^\circ$, the
features of the $P_{2,2}$ probability map show strong similarities 
to the ideal-case result of Fig. 3.

The parameter values used in
our numerical simulations are given in 
Table I (see Appendix), and the methodological details can be
found in Appendix \ref{app:gtensorsimulation}.
In the example discussed below,
the angle enclosed by the dc effective magnetic fields
$\vec{\mathcal{B}}_{L}$ and $\vec{\mathcal{B}}_{R}$ is $\beta = 32^\circ$,
and the Rabi-frequencies (i.e., the amplitudes of the ac effective
magnetic fields) are set to the same value in the two QDs. 
The pulse sequence considered in the simulations
is shown in Fig. 4a. 
To achieve different spin-rotation angles $\theta_L$ and $\theta_R$
 in the two QDs,
different Rabi-pulse lengths, $t_{{\rm burst},L}$ and
$t_{{\rm burst}, R}$, are applied.
In Fig. 4a, $t_{\rm burst,2\pi}$ 
denotes the pulse length corresponding to a $2\pi$ spin rotation.
The Rabi pulses are started simultaneously
on the two QDs, and their lengths are adjusted to the
desired spin rotation angles $\theta_D$ according to $t_{{\rm burst},D} = 
t_{\rm burst,2\pi} \theta_D / 2\pi$.
The $\epsilon$-sweep towards the charge measurement point is
started simultaneously on the two QDs,
once the time $t_{\rm burst,2\pi} + t_{\rm wait}$ elapsed after
the switch-on moment of the Rabi pulses.

Figure 4b shows the  $P_{2,2}$ map resulting from the numerical
simulation, for
the parameter values given in Table \ref{table:param} and 
$t_{\rm wait} = 23$ ps, when the asymmetry is significant.
Deviations from the ideal-case result of Fig. 3, i.e.,
an enhanced [a suppressed] 
$P_{2,2}$ around $(\theta_L,\theta_R) = (3\pi/2,\pi/2)$
[around $(\theta_L,\theta_R) = (\pi/2,3\pi/2)$] are relatively
small, though clearly visible.

As mentioned above, the $P_{2,2}$ probability map depends
on $t_{\rm wait}$ as the Larmor-precession axes of the two spins
are different. Deviations of the $P_{2,2}$ map
from the ideal-case results can be reduced by 
averaging the probability map for $t_{\rm wait}$ in a single
Larmor period.
Figure 4c shows such a $t_{\rm wait}$-averaged 
$P_{2,2}$ map which
is obtained numerically using the same parameters as for 4b,
but averaged for
$t_{\rm wait} \in [0\, {\rm ps},42\, {\rm ps}]$.
The qualitative features of this result are
the same as those of the ideal-case result (Fig. 3);
even the mirror symmetry of the latter 
with respect to the $\theta_L = \theta_R$ diagonal line is retained. 

Performing the simulation for smaller $\beta$ values, the $P_{2,2}$ map approaches the result of the idealized $\hat g_L=\hat g_R$ case. Therefore the angle $\beta$ should be  minimized by choosing an optimized B-field orientation within the range allowed by the requirement of equal Zeeman splittings. For the InAs NW double QD of Ref.~\onlinecite{SchroerPRL2011}, $\beta$ can be tuned below $4$ degrees. In this case, the expected result $P_{2,2}$ is almost identical to the ideal case shown in Fig.~3.
Note that since the g-tensor in a NW QD strongly depends on the	electrostatic confinement potential defining the dot\cite{CsonkaNanoLett2008,NadjPergeNature2010,SchroerPRL2011}, the former can be tuned \emph{in situ} by reshaping the latter by tuning the gate voltages. This can be a helpful feature for optimizing the effective Zeeman fields in the two dots, i.e., to achieve equal Zeeman splittings and parallel effective B-fields.

We conclude that the proposed method could work even if the two QDs have different and anisotropic g-tensors. If the two Zeeman splittings can be tuned equal, and $\beta \lesssim 30$ degree, then the singlet character of the Cooper-pair is reflected in the measured $P_{2,2}(\theta_L,\theta_R)$, similar to the ideal case. Based on the available experimental data on NW QDs \cite{SchroerPRL2011}, these conditions can be fulfilled.

We note that the anisotropy of  the g-tensor might also serve as a resource in identifying the spin state of the split Cooper-pair. By varying the direction of $\vec B_{\rm ext}$ along the intersection of the surfaces associated to the two g-tensors (see Appendix \ref{app:gtensoranisotropic}), the value of the angle $\beta$ enclosed by the local effective fields can be varied.
By optimizing the relative orientation of the two g-tensors (e.g. by defining the QDs in a bent carbon nanotube \cite{FlensbergPRB2010,LaiArXiv2012}), the range in which $\beta$ can be varied can be maximized. The \emph{in situ} tunability of $\beta$ with confinement gates and varying the direction of the external field $\vec B_{\rm ext}$ suggests the possibility of Bell-type tests or tomography of the spin state of individual split Cooper-pairs. A related idea of a Bell-type test based on dc transport was explored recently in detail  by Braunecker et al. \cite{BrauneckerArXiv2013} .

\subsection{Adiabaticity}
\label{sec:disc3}

As discussed in Sec.~\ref{sec:proposal}, the purpose of the proposed experiment demands that the sweep of the on-site energy $\epsilon$ between the preparation point ($\epsilon \approx -U/2$) and the measurement point ($\epsilon < -U$, see Fig.~\ref{fig2}a) should be adiabatic: a $S(1,1)$ initial state in the preparation point should evolve during the sweep along the lower branch of the $S(1,1) - (2,2)$ anticrossing in Fig.~\ref{fig2}a, and end up in the (2,2) state when $\epsilon$ arrives to the measurement point.

Assuming a constant sweep rate $\alpha = \frac{d \epsilon}{d  t}$, the probability $P_d$ of the diabatic [$S(1,1) \mapsto (2,2)$] transition at the anticrossing $\epsilon = -U$ can be approximated by the Landau-Zener formula \cite{Landau-landauzener, ZenerPRSLA1932}:
\bnen P_d = e^{-\frac{2\pi |\tilde \Delta|^2}{\hbar \alpha}}. \eden
To keep $P_d$ below a certain small threshold $P_d^{\rm max} \ll 1$, the sweep rate $\alpha$ should be kept below
\bnen \alpha^{\rm max}  = \frac{2\pi |\tilde \Delta|^2}{\hbar (- \log P_d^{\rm max})}. \eden
Denoting the distance between the preparation and measurement points by $\Delta \epsilon$, the shortest time period $t_{sweep}^{\rm min}$ to meet the required threshold $P_d^{\rm max}$ can be estimated as
\bnen t^{\rm min}_{sweep} \approx \frac{\Delta \epsilon \,  \hbar (-\log P_d^{\rm max})}{2\pi |\tilde \Delta|^2} \eden
For $\tilde \Delta = 50 \mu$eV, sweep range $\Delta \epsilon = 10 \tilde \Delta$, diabatic transition probability threshold $P_d^{\rm max} = 0.1$, we find $t_{sweep}^{\rm min} \approx 50$ ps. A sweep time longer than $t_{sweep}^{\rm min}$ implies smaller diabatic transition probability than $P_d^{\rm max}$.

In the presence of nuclear spins or different g-tensors on the two QDs, an anticrossing might open at the level crossing of $T_-(1,1)$ and the low-energy hybrid state formed by $S(1,1)$ and $(2,2)$. We refer to the value of $\epsilon$ corresponding to this level crossing as $\epsilon_-$ (see $\epsilon_-$ at the x axis of Fig.~\ref{fig2}a). If the charge measurement is carried out at a point $\epsilon < \epsilon_-$, as shown in Fig.~\ref{fig2}a, then it is required to pass through the anticrossing at $\epsilon = \epsilon_-$ \emph{diabatically} during the gate voltage sweep. This requirement together with an expected minimal diabatic transition probability $P_d^{\rm min}\approx 1$ imposes an explicit lower bound $\alpha^{\rm min}$ on the sweep rate $\alpha$ via the Landau-Zener formula. If a time-independent sweep rate is applied between the preparation and manipulation points, then it has to fulfill both requirements, which is possible only if $\alpha^{\rm max} > \alpha^{\rm min}$. In terms of the size of the Hamiltonian matrix  element $\delta$ causing the anticrossing at $\epsilon_-$, the latter condition translates to
\bnen \label{eq:deltacondition} |\delta| < |\tilde \Delta| \sqrt{\frac{\log P_d^{\rm min}}{\log P_d^{\rm max}}}. \eden
Note that this requirement is stronger than $|\delta| < |\tilde \Delta|$. Alternatively, `tailored' gate voltage pulses with time-dependent sweep rates\cite{PettaScience2005,Ribeiro-tailoredpulses} might also be used, or, if the Zeeman splitting exceeds $\tilde \Delta$, the charge measurement can be carried out at an $\epsilon$ between the two anticrossings, $\epsilon_- < \epsilon < -U$.

Even for a relatively large angle $\beta = \pi/6$, the condition \eqref{eq:deltacondition} can be fulfilled. To demonstrate this with a numerical example, we set the diabatic transition thresholds to $P_{d}^{\rm min} = 0.9$ and to $P_d^{\rm max}= 0.1$. With these choices, Eq.~\eqref{eq:deltacondition} translates to $|\delta| < 0.21 |\tilde \Delta|$. Consider the case of $|\hat g_D \mu_B B_{\rm ext}| > \tilde \Delta$, which ensures that the matrix element opening the anticrossing at $\epsilon_-$ is well approximated by the matrix element between $(2,2)$ and the ground state of the $(1,1)$ sector. The latter matrix element is $\delta = \tilde \Delta \sin(\beta/2)/\sqrt{2}$, as can be shown within the framework
outlined in Sec.~\ref{sec:themodel}, after incorporating the effect of different anisotropic g-tensors in Eq.~\eqref{eq:spinhamiltonian}. In the case $\beta = \pi/6$, this is $\delta \approx 0.18 \tilde \Delta$. This fulfills the above requirement, ensuring the possibility to use a constant sweep rate between the preparation and the measurement points and still respect both diabatic probability thresholds.

Note that the above discussion on the gate voltage sweep process is based on a simplified model of two independent Landau-Zener processes. We think that this approach is reliable if either $|\delta| \ll |\tilde \Delta|$ or if the two anticrossings are well separated along the $\epsilon$ axis, i.e., if $|\epsilon_- + U| \ll |\tilde \Delta|, |\delta|$.

\section{Conclusion}

A novel detection method is proposed to demonstrate the singlet character of individual split Cooper-pairs. The QDs coupled to the SC lead are used as detector of the spin character. First the spin state of the electrons is prepared by EDSR technique in the $(1,1)$ charge configuration of the QDs, and then a Cooper-pair is tried to be extracted from the SC adiabatically.  The Pauli principle sets a constraint whether the system could evolve to the $(2,2)$ charge configuration. By measuring the probability of finding the system in the $(2,2)$ configuration at the end of the procedure for different initial spin settings, signature of the singlet character of split Cooper-pair can be demonstrated.

The effect of material parameters were also discussed.  It was shown that the proposed experiment can be also carried out in case of strong  g-factor anisotropy of the QDs, if the effective magnetic field can be set to the same absolute value on the two QDs. However, the presence of strong hyperfine interaction does not allow to demonstrate the singlet character.  The ingredients of the detection method, like the required device geometry, the steps of the manipulation scheme, or the way of the measurement were all demonstrated before, which makes the realization of the proposal feasible with state-of-the-art experimental techniques.

\acknowledgments

We acknowledge useful discussions with J\'anos Asb\'oth, Andreas Baumgartner, Bernd Braunecker, Guido Burkard, P\'eter Domokos, Karsten Flensberg, Ferdinand Kummeth, Zolt\'an Kurucz, Edward Laird, Martin Leijnse, P\'eter Makk, Charles Marcus, Pascu Moca, Vlad Pribiag, Christian Schoenenberger and Gergely Zar\'and.

We acknowledge support from EU ERC CooPairEnt 258789, FP7 SE2ND 271554, the EU Marie Curie grant CIG-293834, the EU GEOMDISS project and Hungarian grants OTKA CNK80991 and OTKA PD 100373. A.~P.~and S.~C.~are supported by the Bolyai Scholarship.

\appendix

\section{Derivation of the effective Hamiltonians}
\label{sec:derivation}

\subsection{The Hamiltonian}

The system considered in the main text consists of a BCS superconductor (SC), and two quantum dots (QDs), $L$ and $R$, which are both tunnel coupled to SC. The corresponding Hamiltonian in the absence of any magnetic field reads
\bnen H = H_L+ H_R + H_S +H_t. \eden

Here the quantum dots are modeled by ($D \in \{L,R\}$)
\bnen H_D = \epsilon n_D + \frac{U}{2} n_D (n_D-1), \eden
where $\epsilon$ is the on-site energy that is assumed to be identical on the two QDs,  $U$ is the on-site Coulomb energy, and $n_D$ is the electron number operator on dot $D$. We assume that the orbital level spacing $\Delta E$ on the QDs is large, therefore the orbital levels lying above the ground-state one are disregarded. Therefore, in our simple model the maximum number of electrons per dot is two. Tunnel coupling as well as capacitive coupling between the two dots are disregarded.

The superconductor is modeled by the BCS Hamiltonian \cite{Schrieffer}:
\bean H_S  &=& \sum_{\vec k s} \xi_{k} c_{\vec k s}^\dag c_{\vec k s} + \sum_{\vec k} \left(\Delta c_{\vec k \uparrow}^\dag c_{-\vec k \downarrow}^\dag + \mbox{h.~c.} \right), \\ &=& \sum_{\vec k s} E_k \gamma_{\vec k s}^\dag \gamma_{\vec k s}, \eean
where $\xi_k$ is the dispersion relation of the electrons in the superconductor in the absence of superconductivity, $\Delta$ is the superconducting gap, $c_{\vec k s}^\dag$ ($c_{\vec k s}$) is an operator creating (annihilating) an electron in the superconductor with wave number $\vec k$ and spin quantum number $s$, $E_k = \sqrt{\xi_k^2+\Delta^2}$ is the dispersion relation of the quasiparticles and $\gamma_{\vec k s}^\dag$ ($\gamma_{\vec k s}$) is an operator creating (annihilating) a quasiparticle. The connection between the electron ($c$) and quasiparticle ($\gamma$) operators is
\bnen \label{eq:gamma} \spinor{c_{\vec k\uparrow}}{c^\dag_{-\vec k \downarrow}} = \ttmatrix{u_k^*}{v_k}{-v^*_k}{u_k}  \spinor{\gamma_{\vec k \uparrow}}{\gamma^\dag_{-\vec k\downarrow}}. \eden
A further useful relation follows from Eq. \eqref{eq:gamma}:
\bnen c_{\vec k s} = u^*_k \gamma_{\vec k s} +s v_k \gamma^\dag_{-\vec k,-s}. \eden
We disregard the phase of the superconducting order parameter, hence we have
\bean u_k &=&  \frac 1 {\sqrt 2} \sqrt{1+ \xi_k/E_k}, \\ v_k &=&  -\frac 1 {\sqrt 2} \sqrt{1- \xi_k/E_k}. \eean

Tunneling processes between the QDs and the superconductor are assumed to be spin conserving and equal for the two dots. Tunneling between SC and the QD $D$ is assumed to be restricted to the single spatial point $\vec r_D$ of the superconductor. Hence the tunneling Hamiltonian reads:
\bnen  \label{eq:Htunnel} H_{t} =t \sum_{D \vec k s} \left(d^\dag_{D s} \psi_s(\vec r_D) + \mbox{h.~c.} \right), \eden
where $\psi_s(\vec r) = \sum_{\vec k} e^{i\vec k \vec r} c_{\vec k s}$.

\subsection{Effective Hamiltonian at $\epsilon \approx 0$}

We use a perturbative approach to determine the relevant part of the energy spectrum of the considered SC-DQD  hybrid system.  We assume that the tunnel coupling $t$ between SC and the QDs is weak, i.e., smaller than the superconducting gap and the on-site Coulomb energy on the QDs:
\bnen t \ll \Delta, U. \eden
Hence we can separate the Hamiltonian $H$ to an `unperturbed' part $H_0 = H_S + H_L+H_R$, and treat the tunneling as a perturbation $H' = H_t$.

The measurement protocol described in the main text makes use of the low-energy electronic states that consist of an even number of electrons in the two QDs and zero quasiparticles in the SC. At $\epsilon \approx 0$, these states are
\bean
\ket{(0,0)}  &=&  \ket{0} ,\\
\ket{S(1,1)}  &=&  \frac 1 {\sqrt 2}  \left( d^\dag_{L\uparrow} d^\dag_{R\downarrow}  -  d^\dag_{L\downarrow} d^\dag_{R\uparrow} \right) \ket{0},\\
\ket{T_+(1,1)}  &=&  d^\dag_{L\uparrow} d^\dag_{R\uparrow} \ket{0},\\
\ket{T_0(1,1)}  &=&  \frac 1 {\sqrt 2}  \left(d^\dag_{L\uparrow} d^\dag_{R\downarrow}  +  d^\dag_{L\downarrow} d^\dag_{R\uparrow} \right)  \ket{0},\\
\ket{T_-(1,1)}  &=&  d^\dag_{L\downarrow} d^\dag_{R\downarrow} \ket{0}, \eean
where the two numbers refer to the electron occupation of the $L$ and $R$ dots, respectively, the preceding label  ($S$, $T_+$, $T_0$, $T_-$) refers  to the spin state in the case of the two-electron states, and
$\ket{0}$ denotes the state in which the electron occupancies of both QDs are zero and the superconductor is in its BCS ground state. States with more electrons as well as states in the $(0,2)$ and $(2,0)$ charge configurations are far above in energy due to the large Coulomb repulsion $U$. States with finite quasiparticle occupation are at an energy distance $\Delta$ above the five relevant $(0,0)$ and $(1,1)$ states.

Restricting the unperturbed Hamiltonian $H_0$ to the five-dimensional relevant subspace, we have
\bnen H_{0r} = 2\epsilon \sum_{\sigma} \ket{\sigma (1,1)}\bra{\sigma (1,1)}, \eden
where $\sigma \in \{S,T_+,T_0,T_-\}$. This Hamiltonian $H_{0r}$ is diagonal in the chosen basis. However, second-order virtual processes mediated by tunneling $H_t$, where each intermediate state consists of a single electron in one QD and a single quasiparticle in the superconductor, induce a weak coupling between the $(0,0)$ and $S(1,1)$ states, as shown below.

Second-order quasi-degenerate perturbation theory  (see, e.g., Appendix B of Ref. \onlinecite{Winkler}) implies that the effective Hamiltonian  representing the above-mentioned second-order virtual processes have the form
\bnen \left[H_{r}^{(2)}\right]_{m,m'} = \frac 1 2 \sum_{l} H'_{ml} H'_{lm'} \left( \frac{1}{E_m-E_l} + \frac{1}{E_{m'}-E_l} \right), \eden
where the summation goes for every eigenstate of $H_0$ that lies outside of the relevant subspace and is coupled to the relevant states via tunneling $H' \equiv H_t$, and $m$ and $m'$ refer to the five relevant states. In our case, the virtual states have the form $\ket{Ds, \vec k s'} = d^\dag_{Ds} \gamma^\dag_{\vec k s'}\ket{0}$.

Straightforward calculation shows that the two nonzero matrix elements of  $H_r^{(2)}$ are $\left[H_r^{(2)}\right]_{S(1,1),(0,0)}$ and $\left[H_r^{(2)}\right]_{(0,0),S(1,1)}=\left[H_r^{(2)}\right]_{S(1,1),(0,0)}^*$, where
\bnen	\left[H_r^{(2)}\right]_{S(1,1),(0,0)} = 2 \sqrt{2} t^2 \sum_{\vec k}	\frac{E_k u_k^* v_k \cos(\vec k \vec \delta)}{\epsilon^2-E_k^2},
\eden
where $\vec \delta = \vec r_L - \vec r_R$ is the relative position of the two points where the electrons tunnel between SC and the QDs. As long as $\epsilon \ll \Delta$, the matrix element can be safely approximated as
\bnen
\left[H_r^{(2)}\right]_{S(1,1),(0,0)} \approx -2 \sqrt{2} t^2 \sum_{\vec k} \frac{u_k^* v_k \cos(\vec k \vec \delta)}{E_k} \equiv \tilde \Delta. \label{eq:tildeDelta} \eden
This result implies that the effective Hamiltonian describing the dynamics of the relevant five-dimensional subspace, including the second-order virtual transitions, have the form shown in Eq.~\eqref{eq:heff0}. The interpretation of this result is straightforward: Cooper-pairs forming the BCS ground state of the superconductor are allowed to co-tunnel out onto the QDs (or the other way around).

The value of the $(0,0)-S(1,1)$ coupling matrix element $\tilde \Delta$ can in principle be evaluated if the electronic dispersion $\xi_k$ in the superconductor is known. The result \eqref{eq:tildeDelta} suggests that the value of $\tilde \Delta$ can be controlled (i) upon fabrication by controlling the distance of the two QDs, and (ii) \emph{in situ} by controlling the S-QD tunneling amplitude $t$ by the voltage on the confinement gate electrodes. We also note that the perturbative approach used here loses its validity if $\tilde \Delta \ll \Delta$ does not hold.

\subsection{Effective Hamiltonian at $\epsilon \approx -U$}

If the QD on-site energy $\epsilon$ is tuned to the vicinity of $-U$, then the five lowest-energy eigenstates of $H_0$ are the $\ket{(2,2)}= d^\dag_{L \uparrow} d^\dag_{L \downarrow} d^\dag_{R\uparrow} d^\dag_{R\downarrow} \ket{0}$ state and the four $(1,1)$ states listed above, all of these having an energy $\approx -2U$. Tunneling $H_t$ induces a perturbative coupling between $(2,2)$ and $S(1,1)$, in the same fashion as explained in the previous section.
In the present case, the virtual intermediate states consist of 3 electrons distributed in the two QDs and a single quasiparticle of the superconductor, which we denote as $\ket{\overline{Ds}, \vec k s'} = d_{Ds} \gamma^\dag_{\vec k s'}\ket{(2,2)}$.

A second-order perturbative calculation analogous to that presented in the previous section, together with the assumption that $|\epsilon -U| \ll \Delta$, yields the effective Hamiltonian shown in Eq.~\eqref{eq:heffmU}.

\section{Nuclear spins}
\label{app:nuclearspins}

In this Appendix, we describe the effect of hyperfine interaction
on the proposed measurement scheme via numerical simulations.
Details of the simulations are provided in Sec. \ref{app:nuclearspinssimulation},
and the results are given and discussed in Sec. \ref{app:nuclearspinsresults}.
The results presented here extend those of Sec. \ref{sec:nuclearspins}.

\subsection{Simulation}
\label{app:nuclearspinssimulation}

In this Appendix, we consider the case of isotropic g-tensors that are 
identical on the two dots; they are characterized by a single g-factor to be 
denoted by $g$. The reference frame is chosen such that
the external magnetic field is applied along the $z$ direction, which also
coincides with the spin quantization axis.
Our simulations are performed in a 6-dimensional Hilbert space
spanned by the states of the (0,0), (1,1) and (2,2) charge configurations.
The basis we use is 
$
	\left\{\ket{(0,0)},
	\ket{\uparrow\uparrow},
	\ket{\uparrow\downarrow},
	\ket{\downarrow\uparrow},
	\ket{\downarrow\downarrow},
	\ket{(2,2)}\right\}.
$

In this basis, the Hamiltonian used in our simulations is expressed as
\begin{widetext}
\label{eq:nuclearspinshamiltonian}
\bnen H =
\left(\begin{matrix} 
0 & 
	0 & 
	-\frac{\tilde \Delta}{\sqrt{2}} & 
	\frac{\tilde\Delta}{\sqrt{2}}
	& 0 
	& 0 
\\
0 & 
	\frac{\mathcal B_L + \mathcal B_R}{2}+2\epsilon(t) & 
	\hbar\Omega_R(t) & 
	\hbar\Omega_L(t) & 
	0 & 
	0
\\
-\frac{\tilde \Delta}{\sqrt{2}} & 
	\hbar\Omega_R(t) & 
	\frac{\mathcal B_L - \mathcal B_R}{2}+2\epsilon(t) & 
	0 & 
	\hbar\Omega_L(t) & 
	\frac{\tilde \Delta}{\sqrt{2}}
\\
\frac{\tilde \Delta}{\sqrt{2}} & 
	\hbar\Omega_L(t) & 
	0 & 
	-\frac{\mathcal B_L - \mathcal B_R}{2}+2\epsilon(t) & 
	\hbar\Omega_R(t) & 
	-\frac{\tilde \Delta}{\sqrt{2}}
\\
0 & 
	0 & 
	\hbar\Omega_L(t) & 
	\hbar\Omega_R(t) & 
	-\frac{\mathcal B_L + \mathcal B_R}{2}+2\epsilon(t) & 
	0
\\ 
0 & 
	0 &
	\frac{\tilde \Delta}{\sqrt{2}} & 
	-\frac{\tilde \Delta}{\sqrt{2}} & 
	0 &
	4\epsilon(t)+2U \end{matrix} 
\right).
\eden
\end{widetext}

In Eq. \eqref{eq:nuclearspinshamiltonian}, 
$\mathcal B_{D} = g \mu_B B_{\rm ext} + \mathcal B_{N,D,z}$ 
($D = L,R$).
This provides an accurate description of the dc effective magnetic field
as long as $g\mu_B B_{\rm ext} \gg \mathcal B_N$, since then
the leading-order expansion of Eq. \eqref{eq:spinhamiltonian}
in the small quantity 
$\mathcal B_N / g\mu_B B_{\rm ext}$
is 
$\vec{\mathcal B}_{D} = g \mu_{B} B_{\rm ext} \hat{\bf{z}} + 
\vec{\mathcal B}_{N,D} \approx g \mu_{B} B_{\rm ext} \hat{\bf{z}} + \mathcal B_{N,D,z}
\hat{\bf{z}}$, with $\hat{\bf{z}}$ being the unit vector pointing in the $z$ direction (i.e., along the external B-field).

The matrix elements proportional to $\tilde \Delta$ are obtained
from Eqs. \eqref{eq:heff0} and \eqref{eq:heffmU} after transforming
from the singlet-triplet basis used in Sec. \ref{sec:themodel} 
to the product-state basis used here.

In the simulation, the on-site energy $\epsilon(t)$ of the QDs is parked at 
$-U/2$ for the spin preparation, and swept linearly in time to the
measurement point at $-3U/2$:
\bnen
\label{eq:epsilont}
\epsilon(t)= 
\begin{cases} 
-\frac{U}{2} & 
	\text{if } 0 \leq t < \frac{2\pi}{\Omega_{Rabi}} \\ 
-\frac{U}{2}-\alpha t & \text{if } \frac{2\pi}{\Omega_{Rabi}}  \leq t < \frac{2\pi}{\Omega_{Rabi}}  + \frac{U}{\alpha} 
\\ -\frac{3U}{2} & \text{if } \frac{2\pi}{\Omega_{Rabi}}  + \frac{U}{\alpha} \leq t. \end{cases}, \eden

The EDSR pulse applied for the spin preparation is assumed
to be on resonance with the Zeeman splitting induced by the external 
magnetic field, and also assumed to 
create an effective ac magnetic field along the $x$ axis,
$\boldsymbol \Omega_{D}(t) =  \Omega_{D}(t) \hat{\bf{x}}$,
and its effect is included in the simulation via 
\bnen 
\label{eq:edsrpulses}
\Omega_{D}(t) = 
\begin{cases} 
\Omega_{Rabi} \cos\left(\frac{g \mu_B B_{ext}}{\hbar}t\right) & \text{if } 0 < t \leq \frac{\theta_D}{\Omega_{Rabi}} \\ 0 & \text{otherwise.} \end{cases} \eden
 and $D=R,L$. 

Numerical values of the parameters used in the simulation are
given in Table \ref{table:param}.

We note that the matrix elements that are proportional to $\tilde \Delta$
in Eq. \eqref{eq:nuclearspinshamiltonian} are $\epsilon$-dependent.
This can be shown by extending the perturbative calculation of Appendix
\ref{sec:derivation} to $\epsilon$ values away from the (0,0)-(1,1) and
(1,1)-(2,2) anticrossings. 
Nevertheless, we disregard this $\epsilon$-dependence in our simulations,
because this extra feature does not lead to qualitative differences in the 
results. 

In our simulations, the initial state is the ground state of the (1,1)
charge sector, i.e., $\ket{\!\downarrow \downarrow}$. 
The time evolution of this initial state, governed by the Hamiltonian
of \eqref{eq:nuclearspinshamiltonian}, is computed numerically
up to $t=t_f\equiv \frac{2\pi}{\Omega_{Rabi}} + \frac{U}{\alpha}$,
cf. Eq. \eqref{eq:epsilont}.
For each run, the $z$ component of the Overhauser field,
$\mathcal B_{N,D,z}$ is assumed to be frozen.
The occupation probability $P_{2,2}$ corresponding to the 
charge measurement is derived from the final state
$\psi(t_f)$ as $P_{2,2} = |\langle (2,2) \ket{\psi(t_f)}|^2$.
The resulting $P_{2,2}$ depends on the values
of the Overhauser fields $B_{N,L,z}$ and $B_{N,R,z}$.
We account for the random nature of the Overhauser fields
by averaging for those assuming a Gaussian distribution 
with standard deviation of $\mathcal {B}_N$, resulting in
\bean
\nonumber
\overline{P}_{2,2} &=& \frac{1}{2\pi \mathcal B_N^2}
\int_{-\infty}^{\infty} d \mathcal{B}_{N,L,z}
\int_{-\infty}^{\infty} d\mathcal{B}_{N,R,z}\\
&\times & e^{-\frac{\mathcal{B}_{N,L,z}^2 + \mathcal{B}_{N,R,z}^2}{2\mathcal{B}_N^2}}
P_{2,2}(B_{N,L,z},B_{N,R,z}).
\eean
We estimate this integral numerically, based on the rectangle 
rule, using a grid for $(\mathcal{B}_{N,L,z},\mathcal{B}_{N,R,z})$ in the range $[-4\mathcal{B}_N,4\mathcal{B}_N]\times [-4\mathcal{B}_N,4\mathcal{B}_N]$ with a resolution of $\mathcal{B}_N/5$ $\times$ $\mathcal{B}_N/5$. 

The $P_{2,2}$ probabilities and the
$\overline{P}_{2,2}$ averages were computed on an $11 \times 11$ 
grid of $(\theta_L,\theta_R)$ in the region $[0,2\pi] \times [0,2\pi]$.
The $\overline{P}_{2,2}$  maps shown in 
Fig. \ref{fig:nuclearspinsresults}  
 are 2D interpolations of  this numerical data.

\begin{table}
\begin{center}
\begin{tabular}{| l | c | c |} 
\hline
name & notation & value 
\\ \hline 
Coulomb energy & $U$ & $4$ meV 
\\ \hline 
Rabi frequency & $\Omega_{Rabi}$ & $2\pi \times 100$  MHz 
\\ \hline 
induced gap & $\tilde\Delta$ & $10 \, \mu$eV
\\ \hline 
external B-field & $| \hat{g} \mu_B \vec{B}_{\rm ext}|$ & $100 \,\mu$eV
\\ \hline 
on-site energy, preparation & $\epsilon_{prep}$ & $-U/2 = -2$ meV 
\\ \hline 
on-site energy, readout & $\epsilon_{readout}$ & $-3U/2=-6$ meV
\\ \hline 
on-site energy sweep rate & $\alpha$ & $\frac{\tilde\Delta^2}{\hbar} = 151.98 \frac{\mu {\rm eV}}{\rm ns}$ 
\\ \hline
\end{tabular}
\caption{Numerical values of the parameters used in the simulations. \label{table:param}}
\end{center}
\end{table}

\subsection{Results}
\label{app:nuclearspinsresults}

\begin{figure}
	\begin{center}
	\includegraphics[width=0.8\columnwidth]{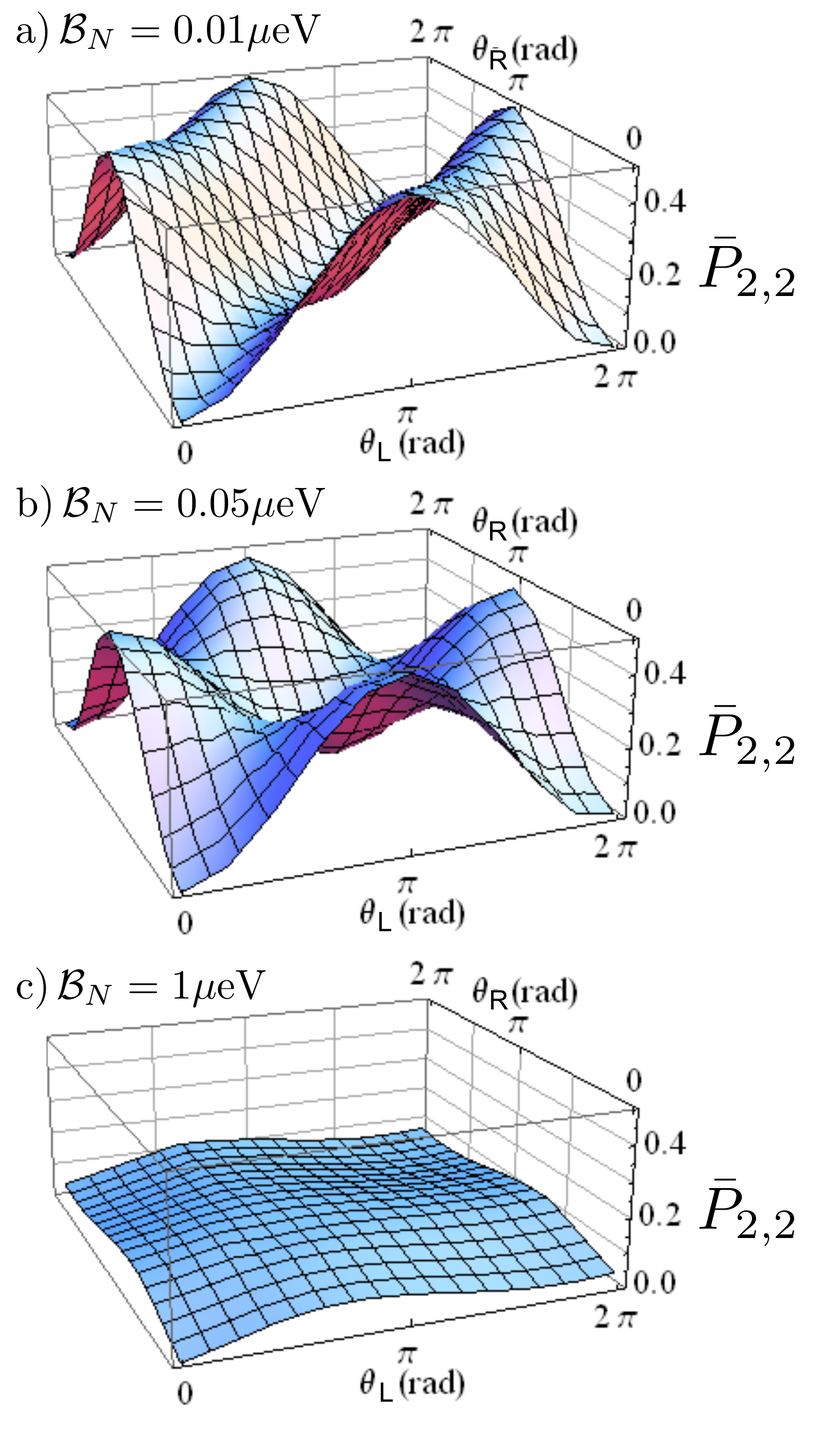}
	\caption{\label{fig:nuclearspinsresults} 
	Deviations from the ideal-case result due to hyperfine interaction.
	The plots show the probability ($P_{2,2}$) of 
	measuring the (2,2) charge state at the end of the
	proposed experimental sequence,
	averaged over the random nuclear-spin configurations 
	($\overline{P}_{2,2}$). 
	Subplots (a), (b) and (c) differ in the energy scale 
	$\mathcal{B}_N$ of the hyperfine interaction. 
	See Appendix \ref{app:nuclearspins} for details 
	and interpretation, and Table \ref{table:param} for the 
	values of the parameters used in the simulation.}
	\label{fig5}
	\end{center}
	\end{figure}

Figure \ref{fig:nuclearspinsresults}a-c show the results for the
Overhauser-field-averaged occupation probability $\overline{P}_{2,2}$
as a function of the spin-rotation angles $\theta_L$ and $\theta_R$,
for three different values of the energy scale $\mathcal B_{N}$ of
the Overhauser fields. 
For comparison, we note that the Rabi frequency $\Omega_{Rabi} = 2\pi \times 100$ MHz we use in the simulations (see Table \ref{table:param})
corresponds to an energy scale of $\hbar \Omega_{Rabi} \approx
0.4 \, \mu$eV.
The key features of the results are as follows. 

(a) In this case, $\hbar \Omega_{Rabi} \gg \mathcal{B}_N$, therefore
the power broadening of the EDSR pulse is large enough to ensure that
the pulse is on resonance with the spins for essentially any value of the 
Overhauser fields.
For the same reason, the hyperfine-induced shift of the spins' Larmor phases
during the spin manipulation,
which are of the order of $\mathcal{B}_N/\hbar \Omega_{Rabi}$, are much
smaller than unity.
These two facts together ensure that $\overline{P}_{2,2}$ shows no qualitative
differences as compared to the ideal-case result (obtained
for the absence of hyperfine interaction) shown in Fig. 3 of the main text.

(b) In this case, the Overhauser-field energy scale $\mathcal{B}_N$ is
still too small to detune the spin splitting from resonance, hence the
spin control is still effective.
This is demonstrated by the feature that the maximal value of $\overline{P}_{2,2}$ approaches 0.5, as in the ideal case (Fig. 3). 
However, the simulation result Fig. \ref{fig:nuclearspinsresults}b  
also demonstrates that the hyperfine-induced
shifts of the spins' Larmor phases, accumulated during the spin manipulation,
are of the order of unity for this parameter set. 
To support this interpretation, let us focus on the special case of 
$\theta_L = \theta_R = \pi/2$, where a saddle point at a height
of $~0.25$ appears in $\overline{P}_{2,2}$ in 
Fig. \ref{fig:nuclearspinsresults}b. 
The interpretation of this value of $\overline{P}_{2,2}$ is as follows. 
The EDSR pulse is effective in creating an equal superposition of the $\uparrow$ and $\downarrow$ states for both spins, implying that 
in the absence of the Overhauser fields, 
the prepared state would have a probability
of 1/4 of occupying both $\ket{T_-(1,1)}$ and $\ket{T_+{1,1}}$,
and a probability of 1/2 of occupying $\ket{T_0(1,1)}$.
However, the Overhauser fields are typically different in the two dots, 
and thereby induce a mixing between $\ket{T_0(1,1)}$ and 
$\ket{S(1,1)}$.
If this mixing is fast enough, which seems to be the case for our 
parameter set, then it results in a 1/4 probability of finding the two spins
in the $\ket{S(1,1)}$ state after the spin preparation.
This interpretation explains the value $P_{2,2}(\pi/2,\pi/2) \approx 0.25$
found in the simulation, and also all further qualitative changes with
respect to the ideal-case result.

(c) In this case, the typical Overhauser field exceeds the power broadening
of the EDSR pulse, i.e, hyperfine interaction detunes the spin splittings
from the resonance condition. 
Therefore the spin manipulation is rendered ineffective, ie, after the pulses
the two-electron spin state remains mostly in the initial state $\ket{\! \downarrow \downarrow}$, leading to a nearly vanishing 
$P_{2,2}$ upon charge measurement.

\section{Different g tensors on the two QDs}
\label{app:gtensor}

In this Appendix, we provide details of the analysis of
the case of different anisotropic g-tensors in the two QDs, 
presented in \ref{sec:disc2}.

\subsection{Anisotropic g-tensors}
\label{app:gtensoranisotropic}


In Sec. \ref{sec:disc2}, we claim that even if the two g-tensors 
$\hat g_L$ and $\hat g_R$
characterizing a DQD are different and anisotropic, it might be possible to 
render the two Zeeman splittings equal by appropriately adjusting the 
direction of the external magnetic field $\vec{B}_{\rm ext}$. 
Here, we provide two examples of such g-tensor pairs, 
see Table \ref{table:gfact}.
Case I  is a hypothetical example, whereas case II is a g-tensor pair
that was measured in a NW DQD\cite{SchroerPRL2011}.

\begin{figure}[!hbp]
	\begin{center}
	\includegraphics[width=\columnwidth]{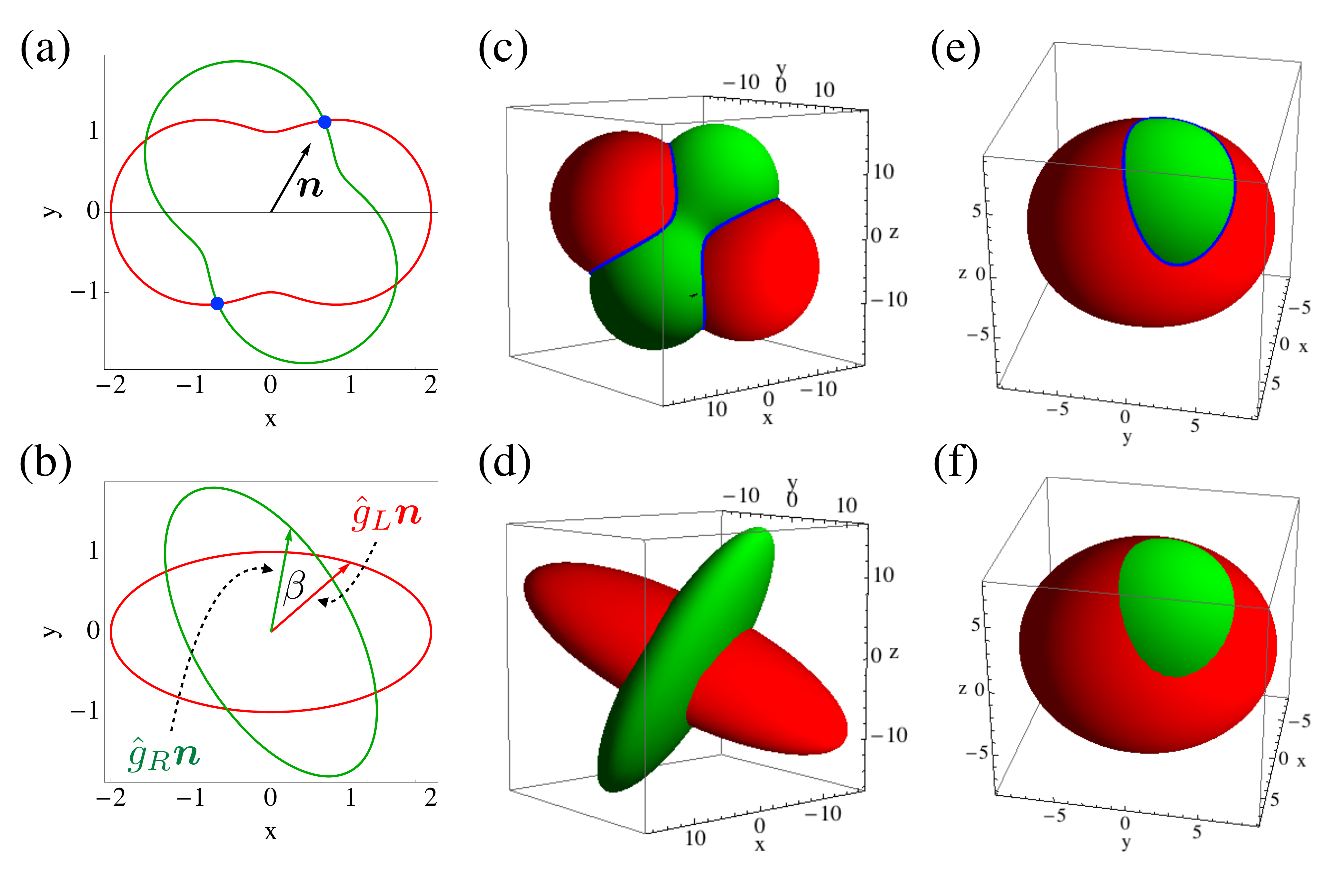}
	\caption{
	Equalizing Zeeman splittings in the two dots. 
	Red/green corresponds to dot $L$/$R$.
	(a) 2D polar plot of the 
	dimensionless Zeeman splitting,
	as a function of the polar angle $\varphi_B$ of the external
	magnetic field.
	Intersections (blue points) correspond to field orientations providing
	equal Zeeman splittings in the two dots. 
	(b) Dimensionless effective magnetic field $\hat{g}_D \vec n$
	corresponding to the field direction $\vec n$ drawn in (a). 
	The ellipses are parametric plots,
	showing $\hat g_D \vec n(\varphi_B)$ as parametrized by 
	$\varphi_B\in[0,2\pi[$.
	(c) 3D spherical plot of the dimensionless Zeeman splitting
	for the g-tensor pair I (see Table \ref{table:gfact}),
	as a function  of the angles $\theta_B$, $\varphi_B$ of the
	external magnetic field.
	Intersections (blue lines) correspond to the field orientations
	providing equal Zeeman splittings in the two dots. 
	(d) The dimensionless effective magnetic field
	$\hat{g}_D \vec n(\theta_B,\varphi_B)$, shown in a 
	parametric plot. 
	(e,f) These correspond to (c,d), 
	with the difference that the g-tensor pair II of Table \ref{table:gfact}
	was used. 
	}
	\label{fig6}
	\end{center}
	\end{figure}

Before discussing cases I and II, let us start with a two-dimensional (2D)
illustration, see Fig. \ref{fig6}a,b. 
We take 
$\hat{g}_L = 
\left(
\begin{array}{cc}
 2 & 0 \\
 0 & 1 \\
\end{array}
\right)
$ and 
$
\hat{g}_R = 
\frac{1}{4} \left(
\begin{array}{cc}
 5 & -\sqrt{3} \\
 -\sqrt{3}& 7\\
\end{array}
\right)
$.
In 2D, the orientation $\vec n = \vec{B_{\rm ext}}/|\vec{B_{\rm ext}}|$ 
of the external magnetic field is 
parametrized with the angle $\varphi_B \in [0,2\pi[$ as 
$\vec n(\varphi_B) = (\cos \varphi_B,\sin \varphi_B)$.
The dimensionless Zeeman splittings are given by
$| \hat {g}_D \vec n |$ ($D =L,R$). 
It is possible to visualize the field orientations of 
equal Zeeman splittings by plotting the 
two dimensionless Zeeman splittings on one 2D polar plot as a function of $\varphi_B$,
see Fig. \ref{fig6}a.
In this figure, the red (green) line corresponds to dot $L$ ($R$). 
The intersection points (blue) of the two lines indicate the field orientations
of equal Zeeman splittings.
One of those field orientations is highlighted in Fig. \ref{fig6}a with
the vector $\vec n$. 
Choosing the field along this $\vec n$, the effective magnetic fields
$\vec{\mathcal B}_D$ enclose a nonzero angle $\beta$ in general.
This is illustrated in Fig. \ref{fig6}b, where
the dimensionless effective magnetic fields $\hat g_{L} \vec n$
and $\hat {g}_R \vec n$ are shown.
(The ellipses in Fig. \ref{fig6}b are parametric plots,
showing $\hat g_D \vec n(\varphi_B)$ as parametrized by 
$\varphi_B\in[0,2\pi[$.)

Now take the three-dimensional (3D) case I in Table \ref{table:gfact}.
The direction of the external magnetic field 
is characterized by spherical coordinates $\theta_B$, $\varphi_B$ fulfilling
\bean \vec n(\theta_B,\varphi_B) 
\equiv
\frac{\vec{B}_{\rm ext} }{| \vec{B}_{\rm ext}|}
= \left( \bna{c} \sin\theta_B \cos \varphi_B \\ \sin\theta_B \sin\varphi_B \\ \cos\theta_B \eda \right). \eean
The field orientations of equal Zeeman splittings are again visualized
by plotting the two dimensionless Zeeman splittings 
on one 3D spherical plot (in analogy with the 2D polar plot above),
see Fig. \ref{fig6}c.
The intersection lines of the two surfaces, shown as blue lines
in Fig. \ref{fig6}c, indicate the field orientations of
equal Zeeman splittings. 
(The corresponding figure for case II is Fig. \ref{fig6}e.)


As demonstrated by the 2D example of Fig. \ref{fig6}a,b,
for equal Zeeman splittings in the QDs, the angle 
$\beta$ enclosed by the effective dc magnetic fields $\vec{\mathcal{B}}_L$
and $\vec{\mathcal{B}}_R$ is generally nonzero. 
In the 3D cases though, see Figs. \ref{fig6}c and e,
the angle $\beta$ can be changed via moving along
the intersection of the two surfaces (i.e., the blue lines), 
while the equality of the Zeeman
splittings is maintained. 
This degree of freedom can, and for the purpose of the 
proposed experiment, should be utilized to minimize $\beta$ with 
the constraint that the Zeeman splittings are equal. 
Our simulations, to be described below, correspond to 
case I (Fig. \ref{fig6}c) 
with such a minimized angle $\beta_{\rm min} \approx 32^\circ$.

\begin{table}
\begin{center}
\begin{tabular}{| c | c || c | c | c | c | c | c |} \hline & QD & $g_1$ & $g_2$ & $g_3$ & $\gamma_1$ & $\gamma_2$ & $\gamma_3$ \\ \hline I & $L$ & 8 & 6 & 12 & 0.9 & 1.1 & -0.75 \\ & $R$ & 5 & 19 & 10 & -0.81 & 2 & 0.5 \\ \hline II &  $L$ & 9.1 & 7.8 & 7.5 & 1.9 & 2.1 & -0.25 \\ &  $R$ & 8.4 & 7.3 & 7 & -0.81 & 2 & 1.5 \\ \hline
\end{tabular}
\caption{The g-tensors for the two examples shown in Fig. \ref{fig6}c-f. 
Here, $g_i$ are the eigenvalues of the g-tensor,
and $\gamma_i$ are the Euler angles, measured in radians, 
characterizing the orientation of the eigenvectors
of the g-tensor.}
\label{table:gfact}
\end{center}
\end{table}

\subsection{Simulation}
\label{app:gtensorsimulation}

\begin{figure}
	\begin{center}
	\includegraphics[width=5cm]{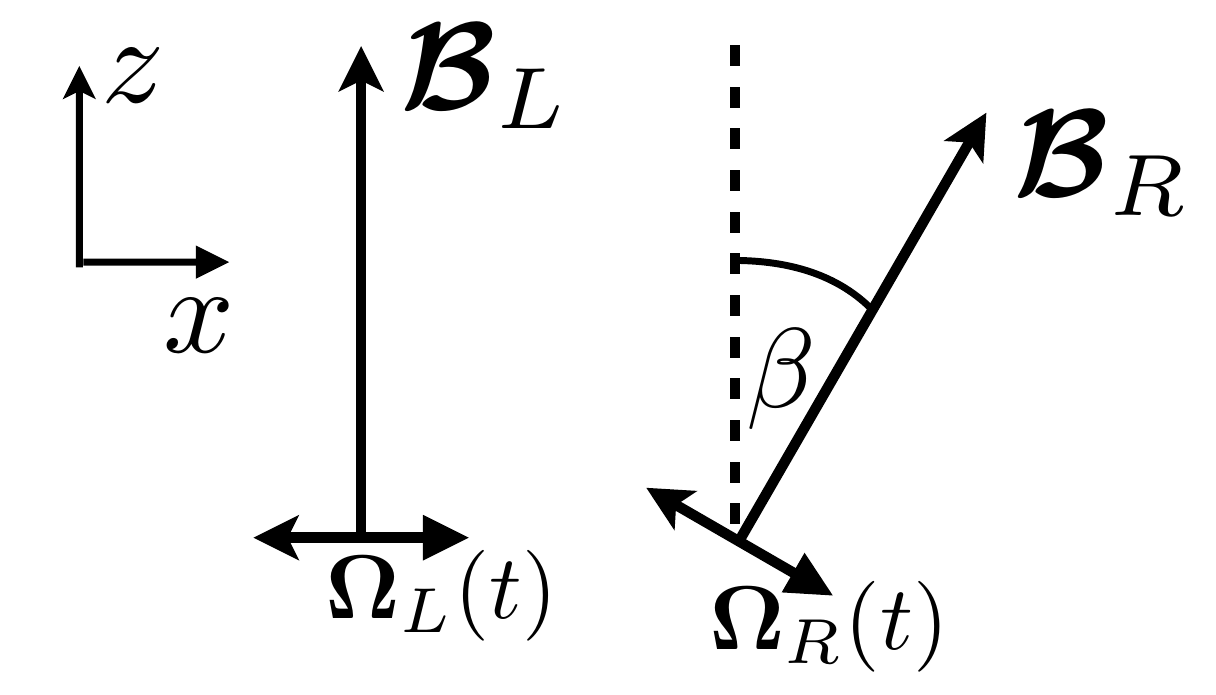}
	\caption{Orientation of the dc ($\vec{\mathcal{B}}_{L}$, 
	$\vec{\mathcal{B}}_{R}$) 
	and ac ($\vec \Omega_L$, $\vec \Omega_R$) 
	effective magnetic fields, as used in the numerical simulations 
	addressing the role of different g-tensors in the two QDs.}
	\label{fig7}
	\end{center}
	\end{figure}

Here, we provide details of the numerical simulations discussed in
Sec. \ref{sec:disc2}.
As stated there, we disregard hyperfine interaction, and
describe a case where the effective dc magnetic fields
felt by the spins, i.e., 
$\vec{\mathcal{B}}_L = \mu_B \hat g_L \vec{B}_{\rm ext}$
and 
$\vec{\mathcal{B}}_{R} = \mu_B \hat g_R \vec{B}_{\rm ext}$,
are equal in magnitude and enclose and
angle $\beta = 32$ degrees.
As illustrated in Fig. \ref{fig7}, we use a reference frame where the $z$ axis is 
aligned with $\vec{\mathcal{B}}_{L}$,
and $\vec{\mathcal{B}}_{R}$ lies in the $x>0$ half-plane
of the $x$-$z$ plane. 
Furthermore, the effective ac fields 
$\vec{\Omega}_{L}$ and $\vec{\Omega}_R$ are assumed to lie
in the $x$-$z$ plane, perpendicular to their respective dc fields.

Our simulations are performed in the 6-dimensional Hilbert space
defined in Sec. \ref{app:nuclearspinssimulation}.
However, here we adjust the basis to our current problem by
using different spin quantization axes for the two dots:
the local quantization axes are aligned with the local dc effective fields.
The corresponding single-spin basis states in QD $L$ ($R$) are denoted by 
$\ket{\uparrow}$ and $\ket{\downarrow}$
($\ket{\Uparrow} \equiv \cos\frac{\beta}{2} \ket{\uparrow}+\sin\frac{\beta}{2} \ket{\downarrow}$ and 
$\ket{\Downarrow}\equiv -\sin\frac{\beta}{2}\ket{\uparrow}+\cos\frac{\beta}{2}\ket{\downarrow}$).
Accordingly, the basis we use here for the 6-dimensional Hilbert space
is $\left\{\ket{(0,0)},\ket{\uparrow\Uparrow},\ket{\uparrow\Downarrow},\ket{\downarrow\Uparrow},\ket{\downarrow\Downarrow},\ket{(2,2)}\right\}$.

In this basis, the Hamiltonian used in our simulations is expressed as
\begin{widetext}
\label{eq:gtensorhamiltonian}
\bnen H =
\left(\begin{matrix} 
	0 & 
	-\frac{\tilde\Delta}{\sqrt{2}}\sin(\frac{\beta}{2}) 	&
	-\frac{\tilde \Delta}{\sqrt{2}}\cos(\frac{\beta}{2}) & 
	\frac{\tilde\Delta}{\sqrt{2}}\cos(\frac{\beta}{2}) & 
	-\frac{\tilde \Delta}{\sqrt{2}}\sin(\frac{\beta}{2}) & 0 
	\\
	-\frac{\tilde \Delta}{\sqrt{2}}\sin(\frac{\beta}{2}) & 
	\mathcal{B}+2\epsilon(t) & 
	\hbar\Omega_R(t) & 
	\hbar\Omega_L(t) & 
	0 & 
	\frac{\tilde \Delta}{\sqrt{2}}\sin(\frac{\beta}{2}) 
	\\
	-\frac{\tilde \Delta}{\sqrt{2}}\cos(\frac{\beta}{2}) &
	\hbar\Omega_R(t) & 
	2\epsilon(t) & 
	0 & 
	\hbar\Omega_L(t) & 
	\frac{\tilde \Delta}{\sqrt{2}}\cos(\frac{\beta}{2}) 
	\\
	\frac{\tilde \Delta}{\sqrt{2}}\cos(\frac{\beta}{2}) & 
	\hbar\Omega_L(t) & 
	0 & 
	2\epsilon(t) & 
	\hbar\Omega_R(t) & 
	-\frac{\tilde \Delta}{\sqrt{2}}\cos(\frac{\beta}{2}) 
	\\
	-\frac{\tilde\Delta}{\sqrt{2}}\sin(\frac{\beta}{2}) & 
	0 & 
	\hbar\Omega_L(t) & 
	\hbar\Omega_R(t) & 
	-\mathcal{B}+2\epsilon(t) & 
	\frac{\tilde \Delta}{\sqrt{2}}\sin(\frac{\beta}{2}) 
	\\ 
	0 & 
	\frac{\tilde \Delta}{\sqrt{2}}\sin(\frac{\beta}{2}) & 
	\frac{\tilde \Delta}{\sqrt{2}}\cos(\frac{\beta}{2}) & 
	-\frac{\tilde \Delta}{\sqrt{2}}\cos(\frac{\beta}{2}) & 
	\frac{\tilde \Delta}{\sqrt{2}}\sin(\frac{\beta}{2}) & 
	4\epsilon(t)+2U 
\end{matrix} \right),
\eden
\end{widetext}
where we introduced $\mathcal{B} = |\vec{\mathcal{B}}_L| = 
|\vec{\mathcal{B}}_R|$.
 
Similarly to the case of Eq. \eqref{eq:nuclearspinshamiltonian},
the matrix elements proportional to $\tilde \Delta$ are 
again obtained from Eqs. \eqref{eq:heff0} and \eqref{eq:heffmU}
after transforming from the singlet-triplet basis used in  
Sec. \ref{sec:themodel} to the product-state basis used in this Section. 
The terms $\propto \tilde \Delta$ in Eq. \eqref{eq:gtensorhamiltonian}
express the fact that for 
a nonzero $\beta$, all four (1,1) states of our current basis
contain a finite amplitude of $\ket{S(1,1)}$, hence the proximity 
effect couples all of them to $\ket{(0,0)}$ and $\ket{(2,2)}$.

As explained in Fig. 4a, in these simulations the on-site energy 
$\epsilon(t)$ of the QDs is parked at $-U/2$ for the spin preparation,
and swept linearly in time to the measurement point at
$-3U/2$:
\bnen 
\label{eq:epsilongtensor}
\epsilon(t)= \begin{cases} -\frac{U}{2} & \text{if } 0 \leq t < \frac{2\pi}{\Omega_{Rabi}} + t_{wait} \\ -\frac{U}{2}-\alpha t & \text{if } \frac{2\pi}{\Omega_{Rabi}} + t_{wait} \leq t < \frac{2\pi}{\Omega_{Rabi}} + t_{wait} + \frac{U}{\alpha} \\ -\frac{3U}{2} & \text{if } \frac{2\pi}{\Omega_{Rabi}} + t_{wait} + \frac{U}{\alpha} \leq t \end{cases}, 
\eden
Note that the difference between this Eq. \eqref{eq:epsilongtensor} and
Eq. \eqref{eq:epsilont} is the appearance of the waiting time $t_{wait}$.

The EDSR pulses $\Omega_D(t)$ used here are identical to 
those given in Eq. \eqref{eq:edsrpulses}.
Numerical values of the parameters used for the simulations 
are given in Table \ref{table:param}.


The $P_{2,2}$ probabilities were computed on a $11 \times 11$ 
grid of $(\theta_L,\theta_R)$ in the region $[0,2\pi] \times [0,2\pi]$.
The $P_{2,2}$ probability maps shown in Figs. \ref{fig4}b,c  
 are 2D interpolations of  the numerical data.



\subsection{Results}
\label{app:gtensorresults}

The results of the numerical simulations are shown in Fig. \ref{fig4}b,c,
and their discussion is included in Sec. \ref{sec:disc2}.

\bibliographystyle{prsty}
\bibliography{scheriff}

\end{document}